\documentclass[twocolumn]{aastex631}
\usepackage{graphicx}
\usepackage{ctable}
\usepackage{amsmath}

\newcommand{\smw}{  \ensuremath{ \sigma_{\mathrm{MW}, i} } }

\newcommand{\sargo}{\ensuremath{ \sigma_{\mathrm{igm},i}^{\mathrm{argo}}}}
\newcommand{\shalo}{\ensuremath{\sigma_{\mathrm{halos},i}}}
\newcommand{\shost}{\ensuremath{\sigma_{\mathrm{host }}^{\mathrm{unk}}}}
\newcommand{\shosth}{\ensuremath{\sigma_{\mathrm{host},i}^{\mathrm{halo}}}}

\newcommand{\rhostar}{\ensuremath{\rho_*}}
\newcommand{\rhob}{\ensuremath{\rho_b}}
\newcommand{\mstar}{\ensuremath{M_*}}
\newcommand{\fdz}{\ensuremath{f_{\rm d}\left( z\right)}}
\newcommand{\fdnoz}{\ensuremath{f_{\rm d}}}

\newcommand{\dmfrb}{\ensuremath{\mathrm{DM}_\mathrm{FRB}}}

\newcommand{\dmhhalo}{ \ensuremath{\mathrm{DM}_{\mathrm{host},i}^{\mathrm{halo}} }}
\newcommand{\dmhunkn}{ \ensuremath{\mathrm{DM}_{\mathrm{host}}^{\mathrm{unk}}  }}
\newcommand{\dmhalos}{\ensuremath{\mathrm{DM}_{\mathrm{halos},i}}}
\newcommand{\dmhalo}{\ensuremath{\mathrm{DM}_{\mathrm{halos}}}}
\newcommand{\dmhost}{\ensuremath{\mathrm{DM}_{\mathrm{host}}}}
\newcommand{\dmigm}{\ensuremath{\mathrm{DM}_\mathrm{igm}}}
\newcommand{\dmigmi}{\ensuremath{\mathrm{DM}_\mathrm{igm,i}}}
\newcommand{\dmargo}{  \ensuremath{\mathrm{DM}_\mathrm{igm}^\mathrm{argo}}}

\newcommand{\mhalo}{\ensuremath{M_{\rm halo}}}

\newcommand{\figm}{\ensuremath{f_\mathrm{igm}}}
\newcommand{\fcgm}{\ensuremath{f_\mathrm{cgm}}}
\newcommand{\fgas}{\ensuremath{f_\mathrm{gas}}}
\newcommand{\ficm}{\ensuremath{f_\mathrm{icm}}}

\newcommand{\fcgmff}{\ensuremath{f_\mathrm{cgm,ff}}}
\newcommand{\fcgmother}{\ensuremath{f_\mathrm{cgm,other}}}

\newcommand{\zfrb}{\ensuremath{z_\mathrm{FRB}}}
\newcommand{\hMpc}{\ensuremath{h^{-1}\,\mathrm{Mpc}}}
\newcommand{\pccm}{\ensuremath{ {\rm pc~cm^{-3}}   }}
\newcommand{\rmax}{\ensuremath{r_{\mathrm{max}}}}

\shorttitle{Inferred Distribution of Baryons}
\shortauthors{Khrykin et al.}

\begin{document}

\title{FLIMFLAM DR1: The First Constraints on the Cosmic Baryon Distribution from 8 FRB sightlines}

\correspondingauthor{Ilya S. Khrykin}
\email{i.khrykin@gmail.com}

\author[0000-0003-0574-7421]{Ilya S. Khrykin}
\affiliation{Instituto de Física, Pontificia Universidad Católica de Valparaíso, Casilla 4059, Valparaíso, Chile}
\affiliation{Kavli IPMU (WPI), UTIAS, The University of Tokyo, Kashiwa, Chiba 277-8583, Japan}

\author[0000-0002-5934-9018]{Metin Ata}
\affiliation{The Oskar Klein Centre, Department of Physics, Stockholm University, AlbaNova University Centre, SE 106 91 Stockholm, Sweden}

\author[0000-0001-9299-5719]{Khee-Gan Lee}
\affiliation{Kavli IPMU (WPI), UTIAS, The University of Tokyo, Kashiwa, Chiba 277-8583, Japan}
\affiliation{Center for Data-Driven Discovery, Kavli IPMU (WPI), UTIAS, The University of Tokyo, Kashiwa, Chiba 277-8583, Japan}

\author[0000-0003-3801-1496]{Sunil Simha}
\affiliation{University of California, Santa Cruz, 1156 High St., Santa Cruz, CA 95064, USA}

\author[0000-0002-0298-8898]{Yuxin Huang}
\affiliation{Kavli IPMU (WPI), UTIAS, The University of Tokyo, Kashiwa, Chiba 277-8583, Japan}

\author[0000-0002-7738-6875]{J. Xavier Prochaska}
\affiliation{University of California, Santa Cruz, 1156 High St., Santa Cruz, CA 95064, USA}
\affiliation{Kavli IPMU (WPI), UTIAS, The University of Tokyo, Kashiwa, Chiba 277-8583, Japan}
\affiliation{Division of Science, National Astronomical Observatory of Japan, 2-21-1 Osawa, Mitaka, Tokyo 181-8588, Japan}

\author[0000-0002-1883-4252]{Nicolas Tejos}
\affiliation{Instituto de Física, Pontificia Universidad Católica de Valparaíso, Casilla 4059, Valparaíso, Chile}

\author[0000-0003-2149-0363]{Keith W. Bannister}
\affiliation{ATNF, CSIRO, Space and Astronomy, PO Box 76, Epping, NSW 1710, Australia}

\author[0000-0001-5703-2108]{Jeff Cooke}
\affiliation{Centre for Astrophysics and Supercomputing, Swinburne University of Technology, Hawthorn, VIC 3122, Australia}
\affiliation{ARC Centre of Excellence for All Sky Astrophysics in 3 Dimensions (ASTRO 3D), Australia}

\author[0000-0002-8101-3027]{Cherie K. Day}
\affiliation{Department of Physics, McGill University, Montreal, Quebec H3A 2T8, Canada}

\author[0000-0001-9434-3837]{Adam Deller}
\affiliation{Centre for Astrophysics and Supercomputing, Swinburne University of Technology, Hawthorn, VIC 3122, Australia}

\author[0000-0002-5067-8894]{Marcin Glowacki}
\affiliation{International Centre for Radio Astronomy Research (ICRAR), Curtin University, Bentley, WA 6102, Australia}

\author[0000-0002-5025-4645]{Alexa C. Gordon}
\affiliation{Center for Interdisciplinary Exploration and Research in Astrophysics (CIERA) and Department of Physics and Astronomy, Northwestern University, Evanston, IL 60208, USA}

\author[0000-0002-6437-6176]{Clancy W. James}
\affiliation{International Centre for Radio Astronomy Research (ICRAR), Curtin University, Bentley, WA 6102, Australia}

\author[0000-0003-1483-0147]{Lachlan Marnoch}
\affiliation{ARC Centre of Excellence for All Sky Astrophysics in 3 Dimensions (ASTRO 3D), Australia}
\affiliation{School of Mathematical and Physical Sciences, Macquarie University, NSW 2109, Australia}
\affiliation{Astrophysics and Space Technologies Research Centre, Macquarie University, Sydney, NSW 2109, Australia}
\affiliation{Australia Telescope National Facility, CSIRO Space and Astronomy, P.O. Box 76, Epping, NSW 1710, Australia}

\author[0000-0002-7285-6348]{Ryan. M. Shannon}
\affiliation{Centre for Astrophysics and Supercomputing, Swinburne University of Technology, Hawthorn, VIC 3122, Australia}

\author[0000-0001-5310-4186]{Jielai Zhang}
\affiliation{Centre for Astrophysics and Supercomputing, Swinburne University of Technology, Hawthorn, VIC 3122, Australia}
\affiliation{ARC Centre of Excellence for Gravitational Wave Discovery (OzGrav), Hawthorn, VIC 3122, Australia}

\author[0009-0002-9608-9275]{Lucas Bernales--Cortes}
\affiliation{Instituto de Física, Pontificia Universidad Católica de Valparaíso, Casilla 4059, Valparaíso, Chile}

\begin{abstract}

The dispersion measure of fast radio bursts (FRBs), arising from the interactions with free electrons along the propagation path, constitutes a unique probe of the cosmic baryon distribution. 
Their constraining power is further enhanced in combination with observations of the foreground large-scale structure and intervening galaxies.
In this work, we present the first constraints on the partition of the cosmic baryons between the intergalactic medium (IGM) and circumgalactic medium (CGM), inferred from the FLIMFLAM spectroscopic survey. In its first data release, the FLIMFLAM survey targeted galaxies in the foreground of $8$ localized FRBs. 
Using Bayesian techniques, we reconstruct the underlying $\sim$Mpc-scale matter density field that is traced by the IGM gas. 
Simultaneously, deeper spectroscopy of intervening foreground galaxies (at impact parameters $b_\perp \lesssim r_{200}$) and the FRB host galaxies constrains the contribution from the CGM. 
Applying Bayesian parameter inference to our data and assuming a fiducial set of priors, 
we infer the IGM cosmic baryon fraction to be $f_{\rm igm}=0.59^{+0.11}_{-0.10}$, and a CGM gas fraction of $f_{\rm gas} = 0.55^{+0.26}_{-0.29}$ for $10^{10}\,M_\odot \lesssim M_{\rm halo}\lesssim 10^{13}\,M_\odot$ halos. The mean FRB host dispersion measure (rest-frame) in our sample is $\langle \mathrm{DM_{host}}\rangle = 90^{+29}_{-19},\mathrm{pc\; cm^{-3}}$, of which $\langle \dmhunkn \rangle =69^{+28}_{-19}\,\mathrm{pc\; cm^{-3}}$ arises from the host galaxy ISM and/or the FRB progenitor environment. While our current \figm{} and \fgas{} uncertainties are too broad to constrain most galactic feedback models, this result marks the first measurement of the IGM and CGM baryon fractions, as well as the first systematic separation of the FRB host dispersion measure into two components:  arising from the halo and from the inner ISM/FRB engine.

\end{abstract}

\keywords{Missing mass (1068), Intergalactic gas (812), Circumgalactic medium (1879), Cosmic web (330), Redshift surveys (1378)}

\section{Introduction} 
\label{sec:intro}

As the gravitational pull of (primarily) dark matter forms the cosmic web, complex and non-linear astrophysical processes conspire to redistribute the baryonic matter. This occurs especially in the potential wells of dark matter halos, where galaxies form and produce supernovae and active galactic nuclei (AGN). Stellar and AGN explosions and radiation heat the gas, pushing it outwards. These ``feedback'' processes can entirely evacuate the potential wells of baryons, driving them far beyond the halo's virial radius \citep[e.g.][]{sorini2022, ayromlou2023,khrykin2023}.

In parallel, the collapse of matter onto the filamentary structures that comprise the cosmic web is predicted to shock-heat the gas and produce the warm-hot intergalactic medium \citep[WHIM;][]{cen1999} that may be the dominant phase of baryons in the $z \sim 0$ universe \citep[e.g.,][]{nevalainen2015}. This WHIM was introduced originally to ``explain'' the missing baryons problem highlighted by \cite{fukugita1998}: the non-detection of $\sim 40 \%$ of the mass density of baryons, \rhob. Yet, despite sustained attempts over the past 20+ years \citep[e.g.][]{lehner2007, tripp2008, narayanan2009,prochaska2011, tejos2016,nicastro2018, degraaf2019},  
observational evidence for the WHIM has been scarce. As such, the (presumed) dominant phase of baryons in the present-day universe is largely unexplored.

Recently, astronomers have leveraged an unexpected phenomenon to resolve the missing baryons problem: fast radio bursts \citep[FRBs;][]{lorimer2007,petroff2022}. Encoded in the signal of these brief pulses of bright radio emission is the dispersion measure (\dmfrb), the integrated electron density along the sightline weighted by the scale factor $a \equiv 1/(1+z)$. 
Unlike most other probes of the IGM, \dmfrb\ is sensitive to the free electron density with only very tiny dependencies on the gas temperature and metallicity, potentially allowing for straightforward interpretations of the observations.
To the extent that the vast majority of extragalactic baryons have been ionized since the end of cosmic reionization 
\citep[$z\lesssim 6$;][]{gunn1965}, 
the free electrons probed by FRBs represent a promising probe of the cosmic baryons.

Analyzing the first handful of localized FRBs, \cite{macquart2020} confirmed the expectation that \dmfrb\ increases with host redshift. The positive correlation is consistent with the cosmic baryon density, $\Omega_b$, derived from early universe observations \citep[e.g.][]{cooke2018}. While this work (and subsequent FRB observations) have {\it detected} the missing baryons, the scarcity of the data and large variance in the Macquart relation caused by the unknown cosmic structures traversed by the FRB makes it challenging to determine the relative location of this otherwise invisible matter \citep[but see][for first attempts from FRBs alone]{baptista2023}.

Thus motivated, we initiated the FLIMFLAM survey \citep{kglee2022} to cross-correlate the \dmfrb{} with foreground structures and galactic halos. The primary goal of the FLIMFLAM survey is to acquire spectroscopic redshifts of galaxies and that way map their distribution in the foregrounds of well-localized FRBs. 
In addition, using a Bayesian algorithm for matter density reconstructions \citep{ata2015,ata2017,ata2021} allows us to significantly account  for the scatter in the observed \dmfrb{} arising from the variance of the large-scale structures. 
\citet{kglee2022} have illustrated how this technique reduces the impact of cosmic variance by a factor of $\simeq 2-3$, and simultaneously constrains the IGM and CGM baryon fractions with far greater precision than feasible with localized FRBs alone (see also \citealp{simha2020}). Such a measurement would also provide insights into the galaxy formation and evolution models. For example, \citet{khrykin2023} showed that different galaxy or active galactic nuclei (AGN) feedback prescriptions can drastically change the relative distribution of baryons between the circum-halo media of galaxies\footnote{Variously known as the circumgalactic medium (CGM), intra-group medium (IGrM), or intra-cluster medium (ICM) depending on the halo mass.} \textit{vis-a-vis} the low-density IGM outside of halos. 

\begin{deluxetable*}{lccccccc}[t]
\label{tbl:tab_dr1}
\tablecaption{List of FRBs in the FLIMFLAM Data Release 1 as used in this work. From left to right the columns show: the ID of a given FRB, right ascension and declination in equatorial J$2000$ coordinates, spectroscopic redshift, survey or instrument used to acquire wide and narrow-field spectroscopic foreground distribution of galaxies, and overall observed \dmfrb{}, as well as the references for these estimates. }
\tablehead{
            \colhead{ FRB } & 
            \colhead{ R.A.  } &
            \colhead{ Decl.  } &
            \colhead{ Redshift } &
            \colhead{ Wide-field } &
            \colhead{ Narrow-field } &
            \colhead{ $\mathrm{DM_{FRB}}$ } &
            \colhead{ Ref. } \\[-4pt]
            \colhead{} & 
            \colhead{(deg)} &
            \colhead{(deg)} &
            \colhead{} & 
            \colhead{data} & 
            \colhead{data} &
            \colhead{($\mathrm{pc\ cm^{-3}} $)} &
            \colhead{} 
          }
\startdata
    20211127A & $199.8088$ & $-18.8381$ & $0.0469$ & 6dF       & AAT           & $234.83$ & Deller et al. (in prep) \\
    20211212A & $157.3507$ & $+01.3605$ & $0.0713$ & SDSS      & AAT           & $206.00$ & Deller et al. (in prep) \\
    20190608A & $334.0199$ & $-07.8982$ & $0.1178$ & SDSS, 6dF & SDSS, KCWI, MUSE   & $339.50$ & \citet{macquart2020} \\
    20200430A & $229.7066$ & $+12.3761$ & $0.1608$ & SDSS, AAT  & LRIS, DEIMOS, MUSE & $380.10$ & \citet{heintz2020} \\
    20191001A & $323.3516$ & $-54.7477$ & $0.2340$ & AAT        & AAT, MUSE, GMOS-S          & $506.92 $& \citet{bhandari2020} \\
    20190714A & $183.9795$ & $-13.0207$ & $0.2365$ & AAT        & LRIS, DEIMOS, MUSE & $504.70$ & \citet{heintz2020} \\
    20180924B & $326.1052$ & $-40.9000$ & $0.3212$ & AAT        & AAT, MUSE         & $362.40$ & \citet{bannister2019} \\
    20200906A & $053.4956$ & $-14.0833$ & $0.3688$ & AAT        & LRIS, DEIMOS, MUSE & $577.80$ & \citet{bhandari2020}
\enddata
\end{deluxetable*} 

In this work, utilizing the ``FRB foreground mapping" technique, we compare the observed \dmfrb{} of the sample of $8$ FRBs that are part of the first FLIMFLAM data release to predictions of baryon distribution from theoretical models, and obtain the first observational constraints on cosmic baryons residing in the diffuse IGM and CGM gas of virialized halos. We utilize the Bayesian Markov Chain Monte Carlo (MCMC) algorithm that takes into account both observational and modelling uncertainties, allowing one to measure the partition of cosmic baryons with high precision.

This paper is organized as follows. In Section~\ref{sec:data}, we discuss the FLIMFLAM observations and archival data that have been used in this work. We describe the density reconstruction algorithm and its results in Section~\ref{sec:argo}. In Section~\ref{sec:dm_model}, we present and describe each component of the model for the observed FRB dispersion measure. We summarize our statistical algorithm for inferring the model parameters gathering the distribution of cosmic baryons, and present the results of the parameter inference from the MCMC in Section~\ref{sec:mcmc}. We discuss our findings in Section~\ref{sec:results} and conclude in Section~\ref{sec:end}.

Throughout this work, we assume a flat $\Lambda$CDM cosmology with dimensionless Hubble constant $h=0.673$, $\Omega_m = 0.315$, $\Omega_b = 0.046$, $\sigma_8=0.8$, and $n_s = 0.96$, consistent with the latest {\it Planck} results \citep{planck2018}.

\section{Data Sample} 
\label{sec:data}

In this paper, we analyze spectroscopic data obtained in the fields surrounding a sample of 8 localized FRBs as listed in Table~\ref{tbl:tab_dr1}. Generally, we selected FRBs that were (i) localized to a host galaxy with high Probabilistic Association of Transients to their Hosts (PATH) posterior probability ($P(O|x) > 0.95$; c.f.\ \citealt{aggarwal2021}); 
(ii) located in regions of the sky with relatively low dust extinction ($E_{\rm B-V} \lesssim 0.06$); and 
(iii) not believed to have a very large 
($\gg 100~\pccm$)
host contribution to the FRB DM (e.g.\ \citealt{simha2023,kglee2023}). 
Our FRBs are derived from the Commensal Real-time ASKAP Fast Transients (CRAFT) Survey conducted on the Australian Square Kilometre Array Pathfinder (ASKAP) radio telescope. These were then followed up with optical facilities by both the CRAFT and the 
Fast and Fortunate for FRB Follow-up (F$^4$)  collaborations\footnote{\url{https://sites.google.com/ucolick.org/f-4}} in order to identify the host galaxies and their redshift.
At the time of observation (2020-2022), these sightlines listed in Table~\ref{tbl:tab_dr1} represented the majority of known localized FRBs that fulfilled the aforementioned criteria.

\subsection{Wide-field Data}
\label{sec:wide}
The analysis in this paper combines the observed dispersion measures from each FRB with detailed spectroscopic observations of their foreground
galaxies. We will publish a separate paper \citep{huang2024} to describe the data in detail in conjunction with our first data release (DR1), but here we provide a broad overview. 

The overall spectroscopic follow-up effort is dubbed the Foreground Line-of-sight Ionization Measurement From Lightcone AAOmega Mapping (FLIMFLAM) Survey, 
which acknowledges the fact that a large fraction of the spectroscopic observations were carried out using the 2dF-AAOmega multi-object fiber spectrograph on the 3.9m Anglo-Australian Telescope (AAT). The AAOmega data represent the bulk of our ``wide'' survey, which represents the shallowest but foundational tier of our `wedding cake' observational strategy, covering thousands of foreground galaxies over $\sim3$ square degrees for each FRB field. The 3D positions of these galaxies will act as tracers for our density reconstruction of the foreground cosmic web towards individual FRB sightlines. 

To select targets for our AAOmega observations, we typically used publicly available imaging survey catalogs such as those from the Dark Energy Survey  \citep[DES;][]{abbott2021}, the Panoramic Survey Telescope and Rapid Response System \citep[Pan-STARRS;][]{chambers2016}, and the Dark Energy Camera Legacy Surveys \citep[DECaLS;][]{dey2019}. 
In each field, we then defined a magnitude limit to select galaxies for spectroscopy based on the redshift of the FRB; for $\zfrb \lesssim [0.15, 0.25,0.4]$, our nominal selection thresholds are dereddened Kron-magnitudes of $r\leq [19.2,19.4,19.8]$, respectively. We dereddened the magnitudes using the Milky Way dust maps of \citet{schlegel1998}. However, we had some confusion between the magnitude definitions from several of the imaging surveys that were only discovered after the spectroscopy was carried out. Cross-comparison between overlapping regions of different imaging surveys was then used to settle on a consistent magnitude definition. Henceforth, we used Kron magnitudes across all our targets. This generally resulted in small modifications to the overall spectroscopic completeness and radial selection functions.  The one exception is that the effective depth of the FRB 20200430A spectroscopy turned out to be $r\leq 18.6$, i.e.\ significantly shallower than the nominal selection threshold of $r\leq 19.2$ for this field. Therefore, we recalculated each field's radial selection functions to account for shallower observations. 

With AAT/2dF-AAOmega, we can observe $\sim 350$ science targets simultaneously in a field of radius $\sim1$ deg (i.e.,\ 3.1 deg$^2$), but there were typically several thousand targets within our magnitude threshold within each 2dF pointing. We, therefore, designed 5-10 separate fiber plate configurations per field to obtain between $1200-2500$ successful galaxy redshifts around each FRB position. The typical exposure times per galaxy range from $40-60$ minutes depending on observing conditions, magnitude threshold, and dust extinction.

For our lowest-redshift sightlines (FRBs 20211127A, 20211212A, 20190608A, 20200430A), we also included publicly available spectroscopic redshift survey data from the 6dF Galaxy Survey \citep{jones2009} and the New York University Value-Added Galaxy Catalog \citep[NYU-VAGC,][]{blanton2005} derived from the legacy Sloan Digital Sky Survey (SDSS, \citealt{abazajian2009}). 
From these catalogs, we incorporated galaxies within 10 deg from the FRB position into our density reconstructions. The large footprint available from these wide-field surveys allowed us to cover a larger transverse distance than possible with the AAT observations. This allows a more accurate density reconstruction at the low-redshift end, which dominates the path length toward these FRBs.

\subsection{Narrow-field Data}
\label{sec:narrow}

For galaxies that might be directly intersected by the FRB sightlines, it was our desire to reach fainter magnitudes than the $L \sim L^{\star}$ galaxies targeted as large-scale structure tracers in the wide-field data. Early on in FLIMFLAM, we decided to target fainter galaxies down to $r\leq 21.5$ within a $2.5\arcmin$ radius around each FRB, which we dubbed our `narrow-field' sample of galaxies. To obtain better signal-to-noise on these fainter galaxies, the `narrow-field' galaxies were each assigned fibers across 2-3 plate configurations within the same field to boost their total exposure time.

As our survey collaboration took shape, we obtained supplementary observing time on larger telescopes to target even fainter galaxies around our FRB sightlines.
These include the DEIMOS and LRIS spectrographs at the W.M.\ Keck Observatory, GMOS on Gemini-South, and MUSE on UT4 of the Very Large Telescopes (VLT). 
Multi-object slit masks were used with the Keck and Gemini spectrographs to target multiple galaxies within $\lesssim 10'$ of each FRB, corresponding to 1-2 virial radii of typical galaxy halo masses. The typical depth of these observations were $r \leq 22.5$, which corresponds to $L\sim 0.1\,L^{\star}$ for galaxies at $z\sim 0.3$ (c.f.\ Figure~4 in \citealt{heintz2020}). 

In the case of our two lowest redshift FRBs (20211127A and 20211212A), our nominal goal to observe $L\sim 0.1\,L^{\star}$ galaxies corresponds to a comparatively bright magnitude limit of $r \approx 20$, for which the 3.9m AAT was deemed sufficient. 
We therefore observed two plates filled only with `narrow-field' galaxies over each of these two fields --- the public SDSS and/or 6dF survey data were deemed sufficient for the `wide-field' large-scale structure tracer galaxies for these low-redshift FRB fields.

\begin{deluxetable*}{lccccccc}[t]
\label{tbl:tab_argo}
\tablewidth{\textwidth}
\tablecaption{ 
Parameters of the \texttt{ARGO} reconstruction volume for the FRB fields analysed in this work (see Table~\ref{tbl:tab_dr1}). From left ot right, the columns show the ID of a given FRB, comoving distance to the FRB redshift, estimated with Equation~\ref{eq:dcom}, number of reconstruction cells and the corresponding ranges along X, Y and Z directions, survey or instrument used to aquire wide field spectroscopic data, the limiting $r$-band photometric magnitude used to select galaxies and estimate ASF/RSF, the final number of galaxies used in \texttt{ARGO} reconstructions.
}
\tablehead{
\colhead{FRB} & \colhead{$d_{\rm FRB}$} & \multicolumn{2}{c}{ARGO volume parameters} & \colhead{WF Survey} & \colhead{Limiting} & \colhead{\textnumero } \\[-6pt]
\colhead{} & \colhead{}  & \colhead{$N_{\rm x}$/$N_{\rm y}$/$N_{\rm z}$} & \colhead{X; Y/Z ranges} & \colhead{source} & \colhead{$r$-mag} & \colhead{of galaxies} \\[-6pt] 
\colhead{} & \colhead{$(h^{-1}{\rm Mpc})$} & & $(h^{-1}{\rm Mpc})$ & \colhead{} & \colhead{} & \colhead{}
} 
\startdata
20211127I & 136.4  & 100/100/100 & 50-237.5;  \ -93.75-93.75 & 6dF  & 15.60 & 1901 \\
20211212A & 210.1  & 100/100/100 & 50-237.5;  \ -93.75-93.75 & SDSS & 17.77 & 15321 \\
20190608B & 343.1  & 175/100/100 & 50-378.1;  \ -93.75-93.75 & SDSS & 17.77 & 6640 \\
          &        &             &                           & 6dF  & 15.60 & 1273 \\
20200430A & 463.9  & 250/100/100 & 50-518.8;  \ -93.75-93.75 & SDSS & 17.77 & 30579 \\
          &        &             &                           & AAT  & 18.60 & 260 \\
20191001A & 661.5  & 350/100/100 & 50-706.3;  \ -93.75-93.75 & AAT  & 19.40 & 1712 \\
20190714A & 677.4  & 350/100/100 & 50-706.3;  \ -93.75-93.75 & AAT  & 19.40 & 1270 \\
20180924B & 884.2  & 475/100/100 & 50-940.6;  \ -93.75-93.75 & AAT  & 19.80 & 2128\\
20200906A & 1006.1 & 525/100/100 & 50-1034.4; -93.75-93.75   & AAT  & 19.80 & 2186\\
\enddata
\end{deluxetable*}

For all sightlines except FRB~20211127I and FRB~20211212A, we also managed to obtain integral field unit (IFU) spectroscopy within $\sim 0.5-1\,\arcmin$ around the FRB positions using the Keck-II/KCWI and VLT/MUSE instruments, respectively. MUSE Observations were conducted in its ``wide-field adaptive optics" mode (WFM-AO), i.e. covering a $1\times 1$\,arcmin$^2$ field, and each field was integrated for $4800$\,s corresponding to a $5\sigma$ depth of $r\sim 25$. The reduced and flux-calibrated cubes were summed in the spectral direction to produce a white-light image to identify sources. Spectra were extracted at the location of the identified sources. To produce the synthetic photometry for sources without data in existing public surveys, we used the g, r, i SDSS filter transmission curves but manually set transmission to 0 beyond the wavelength coverage of MUSE (4800--9300\,\AA). Furthermore, we set the transmission to 0 between 5800--5960\,\AA\ to account for the blocking filter used to avoid the light from the laser guide stars.  

Finally, for each FRB field we also queried through publically available NASA/IPAC Extragalactic Database (NED) Local Volume Sample \citep[NED-LVS;][]{cook2023}, that contains spectroscopic and photometric information (from GALEX, 2MASS, and AllWISE all-sky surveys) about nearby galaxies at cosmic distances up to $100~{\rm cMpc}$ that might intersect within impact parameters of $b_\perp \leq 2$~cMpc.
This helps to supplement our foreground sample with relatively nearby galaxies that might intersect our FRB sightlines despite being outside the field-of-view of our dedicated observations. 

\section{Density Reconstructions}
\label{sec:argo}

To obtain the underlying density field along the line-of-sight of each individual FRB, we utilize the \texttt{ARGO} numerical algorithm \citep{ata2015,ata2017}, based on works by \cite{kitaura2008}, \cite{jasche2010}, and \cite{kitaura2010}. In what follows, we outline the main properties of the \texttt{ARGO} reconstructions used in this work, and refer the reader to more detailed description given in the aforementioned manuscripts, as well as in \cite{kglee2022}, where the multi-tracer extension is described.

\subsection{\texttt{ARGO} Setup}
\label{sec:argo_setup}

\texttt{ARGO} is a fully Bayesian inference algorithm that applies a Hybrid Monte Carlo technique \citep[HMC;][]{duane1987, neal2011}
to reconstruct the evolved cosmic matter density fields given the observed redshift-space distribution of galaxies on the light-cone. The code depends only on the assumed cosmological and structure formation models, once the galaxy survey' selection functions and galaxy bias are taken into account. In addition, our version of \texttt{ARGO} adopts a prescription from \citet{ata2021}, that allows combining information from multiple individual spectroscopic surveys. 

Before running \texttt{ARGO}, first, for a given FRB field, we set up a rectangular comoving reconstruction volume with cell sizes of $1.875\,\hMpc$, where the X axis is aligned with the line-of-sight direction to the FRB. Each volume contains $N_{\rm y} = N_{\rm z} = 100$ cells along the Y and Z axes, respectively, which represent the dimensions perpendicular to the line-of-sight. Our fields are narrow enough that we can adopt the flat-sky approximation, and assume that the plane of the sky is always perpendicular to the line-of-sight. The number of cells along the X axis is adjusted depending on the comoving distance between the observer and an FRB, given by

\begin{equation}
\label{eq:dcom}
    d_{\rm com} = \frac{c}{H_0} \int_0^{z_{\rm spec}} \frac{{\rm d}z}{\sqrt{\Omega_{\rm M}\left( 1 + z \right)^3 + \Omega_{\rm \lambda}}}.  
\end{equation}

Similar to \citet{kglee2022}, we exclude the first $50~h^{-1}{\rm Mpc}$ along the X axis direction due to decreased \texttt{ARGO} performance at nearby comoving distances, where the lightcone distribution of galaxies becomes very narrow and the reconstructions would be noisy. For the first $50\,\hMpc$ of the path, we simply apply the mean cosmic $\langle \dmigm \rangle$ value.  In addition, we extend the X-axis beyond the line-of-sight position of a given FRB to avoid any potential boundary effects in the reconstructions near the FRB location. We provide a summary of the \texttt{ARGO} reconstruction volumes properties in Table~\ref{tbl:tab_argo}. 

To simplify the ${\rm DM_{igm}}$ estimation at later stages of our analysis, the center of the coordinate system of the \texttt{ARGO} volume is chosen in such a way that an FRB is located at Cartesian coordinates $\bm{p}=\{ {\rm X}_{\rm frb}, {\rm Y}=0, {\rm Z}=0\}$. In order to place an FRB at these coordinates, we need to adopt a transformation between the on-sky and the corresponding Cartesian coordinates, provided by

\begin{align*} 
\label{eq:trans}
    & {\rm X} = d_{\rm com} \cos{\alpha} \cos{\delta},\\ 
    & {\rm Y} = d_{\rm com} \sin{\alpha} \cos{\delta},\\
    & {\rm Z} = d_{\rm com} \sin{\delta},
\end{align*}
where $\alpha,~\delta$ are the right ascension and declination coordinates of the FRB, and $d_{\rm com}$ is given by Equation~\ref{eq:dcom}. However, the resulting FRB coordinate vector is not yet aligned with the aforementioned coordinate system of the \texttt{ARGO} volume. Thus, we further estimate the rotation matrix that is then used to rotate the observed FRB frame to the correct \texttt{ARGO} coordinate system. 

\subsection{Galaxy Sample}
\label{sec:argo_wfd}

As mentioned in Section~\ref{sec:wide}, we adopt the wide-field sample of galaxies as tracers of the underlying matter density field in our \texttt{ARGO} density reconstructions. 
Therefore, the positions of the galaxies in these `wide-field' samples have to be mapped onto the \texttt{ARGO} coordinate system as well. Similar to the discussion in the previous section, we first convert galaxy on-sky coordinates to a Cartesian representation.  Then, we adopt the rotation matrix found for the vector of FRB coordinates and rotate these Cartesian coordinates to match the chosen \texttt{ARGO} coordinate system.

In addition to the coordinates of the galaxies, we supply \texttt{ARGO} with their respective stellar masses \mstar. 
We estimate \mstar\ with the publicly available \texttt{CIGALE} algorithm \citep{boquien2019}, which fits a Spectral Energy Distribution (SED) to the photometric information in multiple bands (see discussion in Section~\ref{sec:data}). For the SDSS and 6dF data, we query public archives to acquire the photometric magnitudes of the galaxies in our FRB fields. SDSS data have their own photometry, while for galaxies in the 6dF survey we used magnitudes from the 2MASS \citep{skrutskie2006} and SuperCOSMOS surveys that were used for target selection in the 6dF survey design \citep{jones2009}. 

Similar to \citet{simha2023}, we initialized the \texttt{CIGALE} algorithm assuming the delayed-exponential star-formation history with no burst population, a synthetic stellar population described in \citet{bruzal2003}, the \citet{chabrier2003} initial mass function, dust attenuation models from \citet{calzetti2001}, and dust emission template from \citet{dale2014}, assuming the AGN fraction of $<20\%$.

\texttt{CIGALE} estimates the mean stellar masses with the corresponding uncertainties. In what follows, we adopt the mean stellar mass estimates from \texttt{CIGALE}. We also summarize the main properties of the wide-field samples that are used in the \texttt{ARGO} reconstructions in Table~\ref{tbl:tab_argo}.

\begin{figure*}
    \centering\includegraphics[width=0.6784\textwidth]{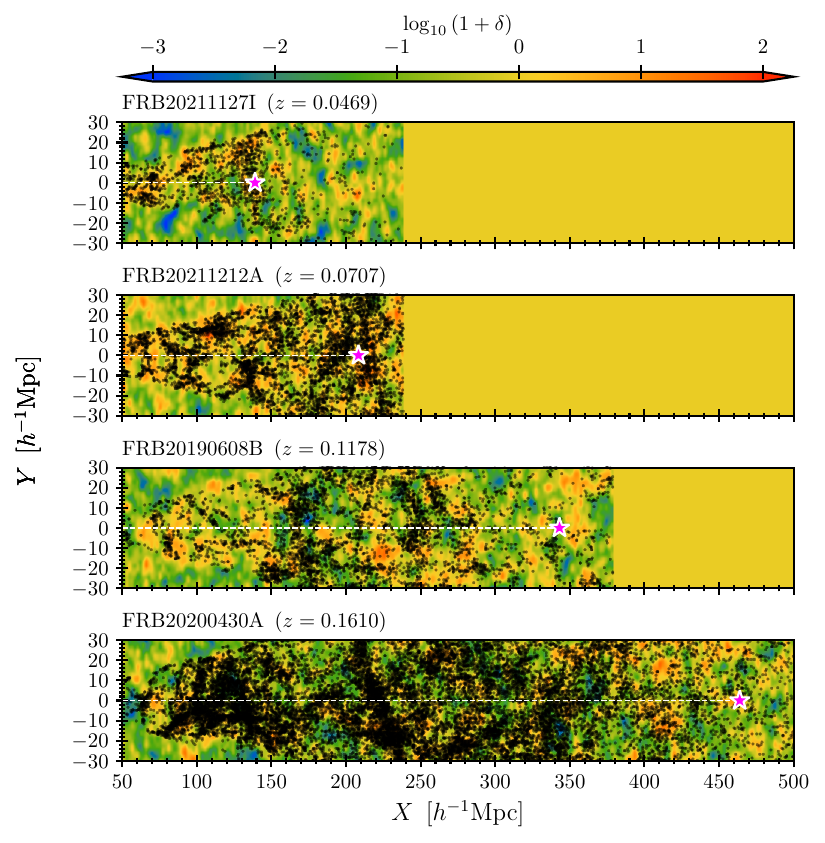}
\caption{ $2$D slices through one realization of the matter density fields reconstructed by the ARGO algorithm. The $X$-coordinate increases with cosmic distance, while the $Y$-coordinate is parallel to the sky plane. The $Z$-coordinate is also parallel to the sky plane, but in this projection we have set it to zero to center the slices on FRB hosts. The black dots illustrate the spatial distribution of the galaxies in the wide- and narrow-fields samples, while the star symbol marks the location of a given FRB host. The dashed line at ${\rm Y}=0$ denotes the line-of-sight of each given FRB. }
\label{fig:figure_recons_p0}
\end{figure*}

\begin{figure*}
    \centering\includegraphics[angle=270,width=0.9\textwidth]{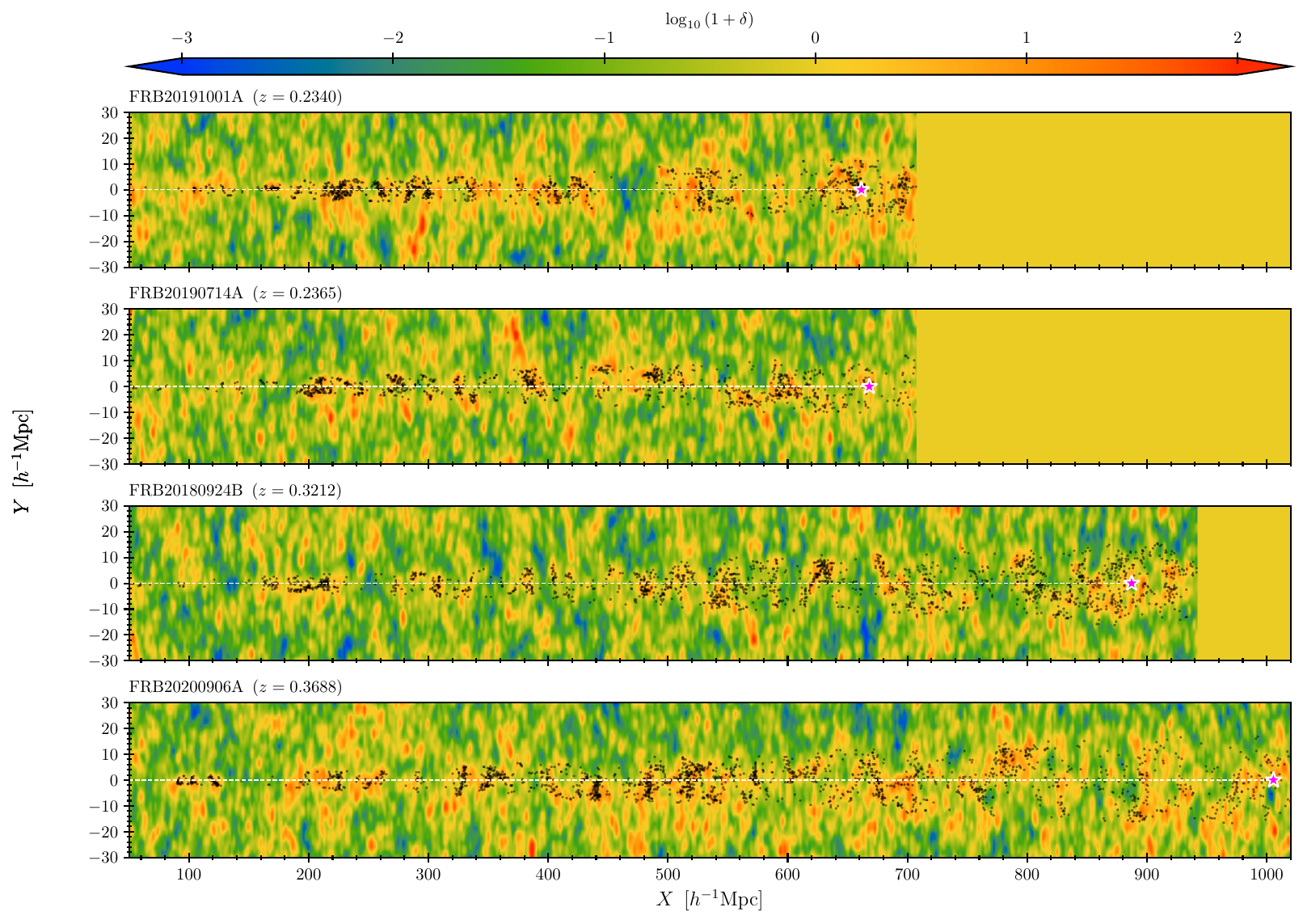}
\caption{Same as Fig.~\ref{fig:figure_recons_p0}, but illustrating the higher redshift FRBs in our sample.}
\label{fig:figure_recons_p1}
\end{figure*}

\subsection{Groups and Clusters of Galaxies}
\label{sec:argo_grps}

\citet{kglee2023} illustrated the importance of taking into account foreground groups and clusters of galaxies when analyzing FRB dispersion measures \citep[see also][]{simha2023}. Omitting this critical information might lead to erroneous conclusions about the nature of the observed DM. Likewise, such large cosmic structures might significantly affect the density reconstructions that do not capture such
massive, non-linear structures. 
Therefore, following the discussion in \citet{kglee2023}, we apply an anisotropic friends-of-friends (FoF) group-finding algorithm \citep[][see also \citealp{tempel2012,tempel2014}]{tago2008} on the galaxy samples in each FRB field. 

This FoF finder adopts a redshift-dependent transverse linking length, $d_\mathrm{LL,\perp}$, given by

\begin{equation}
d_\mathrm{LL,\perp}\left( z\right) = d_\mathrm{LL,0}[1 + a \,\mathrm{arctan}(z/z_*)],
\end{equation}
where $d_\mathrm{LL,0}$ is the linking length at a fiducial redshift, and parameters $a$ and $z_*$ govern the redshift evolution. Such redshift-dependent modification of the linking length allows one to account for the decreasing completeness of the flux-limited spectroscopic surveys. The radial linking length, $d_\mathrm{LL,\parallel}$, is then assumed to by simply proportional to $d_\mathrm{LL,\perp}$. In this work, we adopt the following parameter values: $d_\mathrm{LL,\perp} = 0.35~\hMpc$, $a = 0.75$, $z_* = 0.1$, $d_\mathrm{LL,\parallel}/d_\mathrm{LL,\perp} = 10$.

In order to have a more robust sample, we further refine the FoF algorithm findings by taking into account only the groups with richness parameter $N_{\rm gal} > 5$. Once the galaxy groups are identified in each FRB field, we remove their individual member galaxies from the compiled `wide-field' galaxy samples and replace them with the information (redshift, coordinates, and halo mass) on their corresponding group. The right-hand column of Table~\ref{tbl:tab_argo} shows the final numbers of halos that are derived from the `wide-field' sample of galaxies, serving as input to \texttt{ARGO}.

\subsection{Selection Functions}
\label{sec:sf}

The final ingredient of the \texttt{ARGO} machinery is the information about angular and radial selection functions (ASF/RSF) of the various wide-field surveys, listed in Table~\ref{tbl:tab_dr1}. 
It is crucial to incorporate this information to accurately determine, e.g., whether a given underdensity of galaxies within the survey volume is due to a cosmic void or lack of observations within that region. 
In this section, we briefly outline the basic aspects of the ASF/RSF calculations and refer the interested reader to the detailed description provided by \citet{huang2024}.

For the FRB fields that contain SDSS survey data, we follow the strategy outlined in \citet{ata2021} and extract the ASF in the  $10 \times 10$~deg$^2$ region around the position of the FRB from the publicly available \texttt{MANGLE} outputs\footnote{\url{https://space.mit.edu/~molly/mangle/download/data.html}} \citep{hamilton2004,swanson2008}. 
For the FRB fields containing 6dF survey data, we apply the ASF estimation algorithm in \citet{ata2021}, by comparing the final 6dF DR3 galaxy catalog \citep{jones2009} with the map of 6dF on-sky pointings and stellar masks (communicated privately), again over a $10\times 10$~deg$^2$ field around the FRB. Similarly, the ASF of the AAT survey data is calculated by comparing the number of galaxies with good-quality redshifts to the lists of the selected targets for AAOmega observations combined with the corresponding stellar masks. 

Finally, to estimate the RSF in each field, we compute the distribution of observed galaxies in the \texttt{ARGO} reconstruction volumes as a function of comoving distance from the observer, in bins of $10~\hMpc$. 

\subsection{\texttt{ARGO} Reconstructions Results}
\label{sec:argo_res}

Once all the \texttt{ARGO} inputs are prepared, we initialize the reconstructions in each FRB field and run the HMC sampler for 12\,000 iterations. We find that the posterior samples have a correlation length of $\sim 100-150$, which allows us to extract $N=61$ posterior realizations of each reconstructed density field, separated by at least 150 iterations on the chain. 

In Figures~\ref{fig:figure_recons_p0}~and~\ref{fig:figure_recons_p1}, we show one realization each of the reconstructed matter density field in the foreground of all FRBs in our sample. The density field is smoothed with a Gaussian kernel with the size $R=0.7~\hMpc$, and the corresponding redshift space distribution of the galaxies in the `wide-field' sample is shown by the black dots. 

Because \texttt{ARGO} yields multiple posterior realizations of the reconstructed matter density field per each FRB foreground region, we are able to not only estimate the integrated dispersion measure along the given FRB line-of-sight but also to estimate the corresponding statistical uncertainty of the reconstruction, which we discuss in the next Section. 

\section{Dispersion Measure Model}
\label{sec:dm_model}

One of the key characteristics of any FRB signal is the dispersion measure -- a time delay of arrival of photons at different frequencies. As an integral measure, it is typical to model the observed \dmfrb{} as contributions from several components. For each $i$th FRB in our sample, we assume a model for the observed \dmfrb, given by

\begin{equation}
\label{eq:dm_model}
    {\rm DM}_{{\rm model},i} =  {\rm DM}_{{\rm MW},i} + {\rm DM}_{{\rm cosmic},i} + {\rm DM}_{{\rm host},i},
\end{equation}
where ${\rm DM}_{{\rm MW}}$ represents the contribution from the interstellar medium (ISM) and halo of the Milky Way, ${\rm DM}_{\rm host}$ comes from the FRB host galaxy, while 
\begin{equation}\label{eq:dm_cosmic}
{\rm DM}_{{\rm cosmic},i} = \dmigmi + \dmhalos
\end{equation}
is the contribution from the foreground cosmic structures in the diffuse IGM, \dmigm, and intersected foreground galactic halos, \dmhalo, respectively.

Note that in the FRB literature, the notation \dmigm\ is often used to collectively denote both the IGM and CGM contributions, which we consider separately in Equation~\ref{eq:dm_cosmic} and in our analysis. Roughly speaking, \dmigm\ arises from gas tracing the low-density ($0 \lesssim \rho/\bar{\rho} \lesssim 10$) voids and filaments of the cosmic web, while \dmhalo\ arises from intersections approximately within the virial radii of galactic halos, at matter densities of $\rho/\bar{\rho} \gtrsim 10-100$.

In what follows, we discuss these components separately and describe our model and our adopted parameters for DM estimation.

\subsection{The Milky Way}
\label{sec:dm_mw}

The DM contribution from the Milky Way is given by
\begin{equation}
\label{eq:dmmw}
    {\rm DM}_{{\rm MW},i} = {\rm DM}_{{\rm MW},i}^{\rm ISM} + {\rm DM}_{\rm MW}^{\rm halo},
\end{equation}
where ${\rm DM}_{{\rm MW},i}^{\rm ISM}$ arises from the interstellar medium (ISM), while ${\rm DM}_{\rm MW}^{\rm halo}$ is contributed by the ionized gas in the Milky Way's halo. 

In this work, we adopt ${\rm DM}_{\rm MW}^{\rm ISM}$ values estimated from the \texttt{NE2001} model \citep{cordes2002} based on their Galactic latitude and longitude, which were conveniently tabulated in \citet{james2022b}. The Galactic halo contribution, on the other hand, is given by the estimates of \citet{xyz2019}. In what follows, we adopt the mean value of ${\rm DM}_{\rm MW}^{\rm halo} = 40~\pccm$ for each FRB field, and include an uncertainty of $\sigma_{\rm MW} = 15~\pccm$ into the inference algorithm (see Section~\ref{sec:like}). We list the corresponding combined $\mathrm{DM}_{\rm MW}$ values in each FRB field in Table~\ref{tbl:tab_dms}. 

We note that ${\rm DM}_{\rm MW}^{\rm halo}$ remains a highly uncertain quantity. Previous studies place it in the range $ 10~\pccm \lesssim {\rm DM}_{\rm MW}^{\rm halo} \lesssim 111~\pccm $ \citep{keating2020, coocka2023, ravi2023}. However, the exact choice of the mean and the uncertainty on ${\rm DM}_{\rm MW}^{\rm halo}$ is a sub-dominant error component in our analysis given the FLIMFLAM DR1 limited data sample.

\subsection{The IGM}
\label{sec:dm_igm}

The \dmigm, arising from the low-density intergalactic gas tracing the large-scale cosmic web along the path $s$ to the FRB, is given by 

\begin{equation}
\label{eq:dmigm}
    \dmigm = \int \frac{n_{e, {\rm igm}}(s)}{1+z(s)} \mathrm{d}s,
\end{equation}
where $n_{e,{\rm igm}}$ is the number density of free electrons residing in the IGM along the sightline. For each FRB field in our sample, we estimate \dmigm\ directly from the \texttt{ARGO} density reconstructions, adopting the discretized version of Equation~(\ref{eq:dmigm}) as follows:

\begin{equation}\label{eq:dm_argo}
    \dmigm^{\rm argo} = \Bar{n}_{e,{\rm bar}}\left( \bar{z}\right) \sum_s \left( 1 + \delta_{m,s}^{\rm sm} \right) l_s \left( 1 + z_s \right)^{-1},
\end{equation}
where $l_s$ is the path length to the cell $s$ of the \texttt{ARGO} reconstruction volume along the FRB line-of-sight, $z_s$ is the corresponding redshift of the cell, $\delta_{m,s}^{\rm sm}$ is the smoothed matter overdensity (see example in Figures~\ref{fig:figure_recons_p0}~and~\ref{fig:figure_recons_p1}). The smoothing length is $R=0.7\,\hMpc$, which was found by \citet{kglee2022} to allow dark matter-only N-body simulations with $1.875\,\hMpc$ grid cells to match the global \dmigm{} distribution in cosmological hydrodynamical simulations. We define $\Bar{n}_{e,{\rm bar}}\left( \bar{z}\right) $ as the mean cosmic density of electrons at the median redshift $\bar{z}$ traversed by the ensemble of FRB paths, defined as

\begin{figure}
    \centering
    \includegraphics[width=\columnwidth]{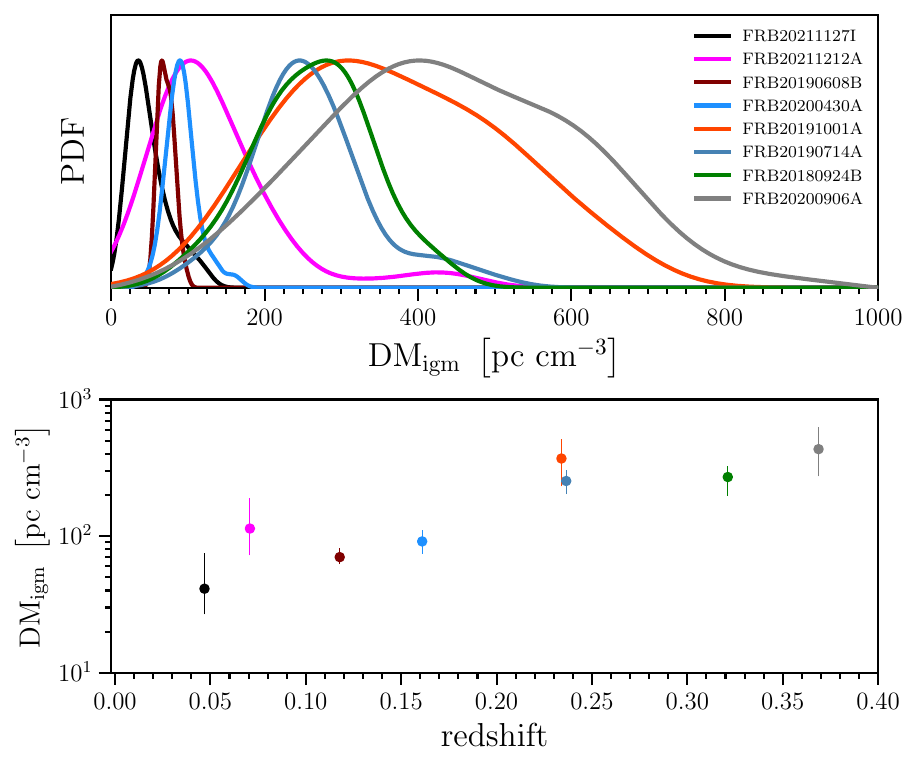}
    \caption{{\it Top:} KDE-smoothed PDF of the ${\rm DM_{\rm igm}}$ distributions estimated for each considered FRB field from the corresponding \texttt{ARGO} reconstructions, with $\figm=0.8$ for illustrative purposes. {\it Bottom:} Estimated values of ${\rm DM_{\rm igm}}$ from \texttt{ARGO} reconstructions as a function of FRB redshift. Each marker is the $50^{\rm th}$ percentile of the corresponding distribution from the top panel, while the error bars are given by $16^{\rm th}$ and $84^{\rm th}$ percentiles of the same distributions.}
    \label{fig:dmi_all}
\end{figure}

\begin{figure*}
    \centering
    \includegraphics[width=0.9\textwidth]{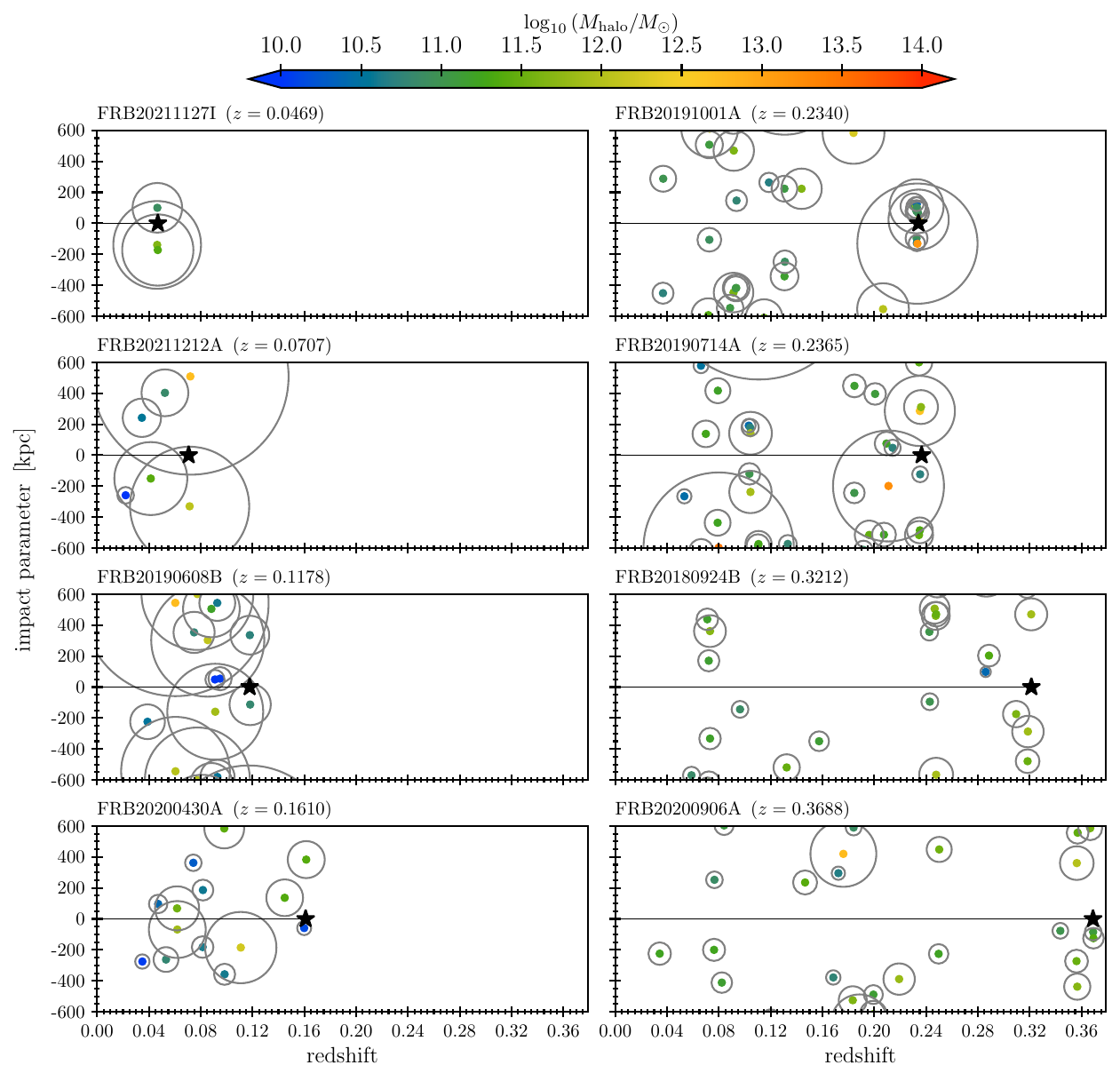}
    \caption{Observed distribution of foreground halos of galaxies and groups around the line-of-sight of the FRBs in the sample. The circles show the estimated size of the halos, $r_{\rm 200}$, given the mean estimate of the corresponding halo masses. The sizes of the halos are only valid along the y-axis; for visualization purposes, the circles representing halo sizes are not to scale along the x-axis. The color of the points represents the halo masses as given by the color-bar.}  
    \label{fig:fg_halos}
\end{figure*}

\begin{equation}\label{eq:n_bar}
    \Bar{n}_{e,{\rm bar}}\left( \bar{z}\right) = \Omega_b \Bar{\rho_c}\left( \bar{z}\right) \left[ \frac{m_{\rm He} \left( 1 - Y\right) + 2Ym_{\rm H}}{m_{\rm He} m_{\rm H}} \right],
\end{equation}
where $m_{\rm H}$ and $m_{\rm He}$ are the atomic masses of hydrogen and helium atoms, respectively; $Y_{\rm He}=0.243$ is the cosmic mass fraction of doubly ionized helium, $\Omega_b=0.044$ is the cosmic baryon density, and $\Bar{\rho_c}\left( \bar{z}\right)$ is the critical density of the Universe. As written, $\Bar{n}_{e,{\rm bar}}$ assumes that all baryons in the Universe are ionized and reside in the IGM. We can further introduce \figm, the fraction of all cosmic baryons residing in the IGM, which will be one of the free parameters in our analysis. Thus,
\begin{equation}
\label{eq:n_e} 
\Bar{n}_{e,{\rm igm}} \equiv \figm \,\Bar{n}_{e,{\rm bar}}
\end{equation}
is the actual mean number density of free electrons in the IGM as constrained by our data. To tie together Equations~\ref{eq:dmigm}, \ref{eq:dm_argo}, \ref{eq:n_bar}, and \ref{eq:n_e} with the current limited data set, we will constrain \figm\ as a free parameter assuming a fixed redshift of $\bar{z}\simeq 0.20$ which is approximately the median redshift probed by the DR1 FRB sightlines. 

\begin{figure}[h]
    \centering
    \includegraphics[width=\columnwidth]{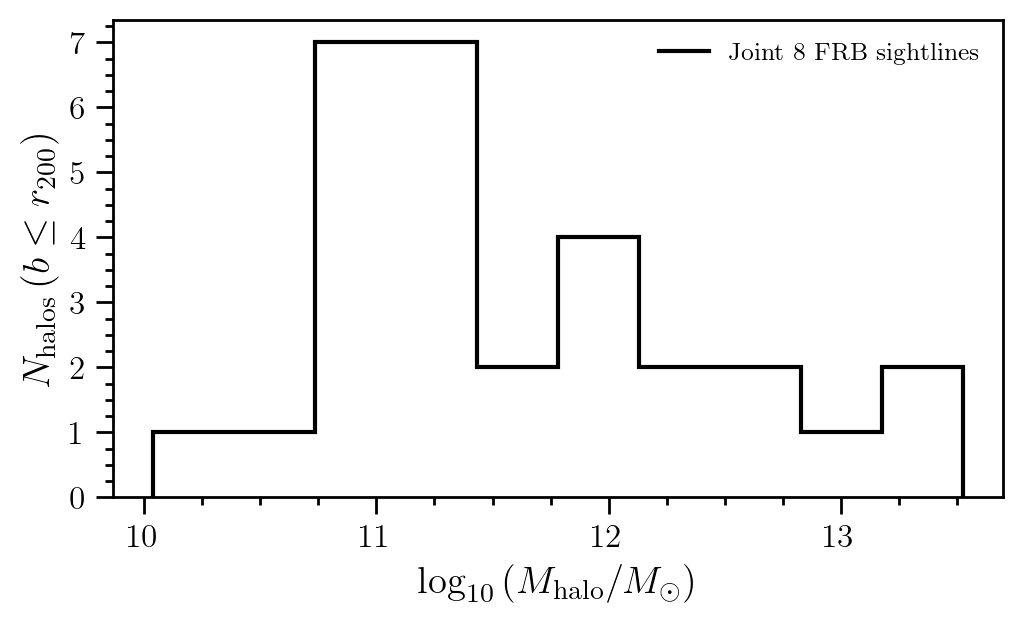}
    \caption{Distribution of the intersected foreground halo masses in the FLIMFLAM DR1 sample. Only halos with impact parameters smaller than the corresponding $r_{200}$ radius (assuming the mean halo mass estimate from CIGALE) of our FRB sightlines are plotted.}  
    \label{fig:hmasses}
\end{figure}

The top panel of Figure~\ref{fig:dmi_all} shows inferred distributions of \dmigm\ in each FRB field in our sample, evaluated from the corresponding $N_{\rm real}=61$ posterior realizations of the matter density field estimated by \texttt{ARGO}. As discussed in Section~\ref{sec:argo_setup}, to take into account the first $50~\hMpc$, excluded from \texttt{ARGO} reconstructions, we add $15~\pccm$ to each individual \dmigm\ posterior value. From each \dmigm\ distribution, we then calculate the mean, $\langle \dmargo \rangle$, and standard deviation $\sigma_{\rm igm}^{\rm argo}$, and plot the results as a function of FRB redshift in the bottom panel of Figure~\ref{fig:dmi_all}. 

We adopt the median of the \texttt{ARGO} realizations for each FRB, scaled by \figm, as the IGM contribution to our model DM in Equation~(\ref{eq:dm_model}), given by 

\begin{equation}
\label{eq:dm_igm_real}
    {\rm DM}_{{\rm igm},i} = \figm \sum_{j}^{N_{\rm real}} \frac{ {\rm DM}^{\rm argo}_{{\rm igm},ij}}{ N_{\rm real} },
\end{equation}
for each FRB sightline $i$.

\subsection{The Intervening Halos}
\label{sec:dm_halo}

To model the contribution of the foreground halos to the observed DM, in each FRB field we adopt the modified Navarro-Frank-White model \citep[mNFW;][]{xyz2019} to estimate the radial density distribution of the gas in each galaxy halo from the corresponding `narrow-field' sample (see Section~\ref{sec:narrow}). In this model, the baryon density is given by

\begin{equation}
\label{eq:mnfw}
  \rho_b \left( r \right) = \fgas \frac{\Omega_b}{\Omega_m} \frac{\rho_0 \left( M_{\rm halo} \right)}{y^{1-\alpha} \left( y_0 + y\right)^{2+\alpha}},
\end{equation}
where \fgas\ is the fraction of cosmic baryons residing in the CGM of each individual galactic halo, relative to the total amount of baryons within the halo if the cosmic baryon fraction were assumed, i.e., $M_{\rm cgm} = \fgas \left( \Omega_b / \Omega_m \right)M_{\rm halo}$;  $\rho_0$ is the central density of the halo as a function of halo mass \mhalo, $y \equiv c \left( r/r_{200}\right)$, $c$ is the concentration parameter, while $y_0$ and $\alpha$ are the mNFW profile parameters. In this work, we use the fiducial values $y_0 = \alpha=2$ from \citet{xyz2019}, and adopt \fgas\ as another free model parameter in our model DM. 

As described in Section~\ref{sec:argo_wfd}, we use \texttt{CIGALE} to estimate the stellar masses of the galaxies in the `narrow-field' samples. We then convert the average stellar masses found by \texttt{CIGALE}  to the corresponding halo masses, $M_{\rm halo}$, by adopting the mean stellar-to-halo mass relation described in \citet{moster2013}. 

Similar to the discussion in Section~\ref{sec:argo_grps}, we use the FoF finder to check if any of the galaxies in the `narrow-field' sample are members of groups and/or clusters. If any groups are found, we remove entries of the member galaxies from our list and add the information on their respective groups.  Figure~\ref{fig:fg_halos} illustrates all foreground halos of galaxies and groups found in the foreground of all FRB sightlines. The circles illustrate the corresponding sizes of the halo' virial radii (only to scale in the vertical, transverse to the line of sight, direction), while the colors indicate the average halo masses inferred from \texttt{CIGALE}.

The overall distribution of halo masses, intersected by the FRB sightlines, is shown in Figure~\ref{fig:hmasses}.
We mostly probe halos with masses of $10^{11}\,M_\odot \lesssim \mhalo \lesssim 10^{12.5}\,M_\odot$, although a small number of our halos are approaching the dwarf galaxy regime ($\mhalo \sim 10^{10}-10^{11}\,M_\odot$) while at the other extreme we probe several galaxy group halos with $\mhalo \sim 10^{13}\,M_\odot$.

Note, that in contrast to \citet{kglee2022} who assumed the mNFW truncation radius $r_{\rm max}$ as a free parameter, we adopt a fixed $\rmax=r_{200}$. While the exact size of the galactic halos is not well-constrained, \citet{simha2020} showed that extending the mNFW profiles to $2\times r_{200}$ can double the \dmhalo\ contribution \citep[see also][]{xyz2019,kglee2023}. 
These results suggest that, in principle, the extent of galactic halos should be included as another free parameter (similar to the discussion in \citealp{kglee2022}). However, the present DR1 sample is of limited constraining power, and the truncated mNFW profile is in any case an approximation. In this work, we, therefore, adopt a fixed truncation radius and leave more refined parametrization to future work
with larger samples.

Finally, to obtain the cumulative \dmhalos\ along the line-of-sight of the $i$th FRB, we sum up the DMs of all individual foreground halos that are found by integrating their respective mNFW profiles in Equation~(\ref{eq:mnfw}). The integration is performed along the intersecting paths of the FRB sightline corresponding to the impact parameters $b_{\perp}$ of the foreground halos, determined by their respective redshifts and angular positions. 

\begin{deluxetable*}{lccccccc}[t]
\label{tbl:tab_dms}
\tablecaption{Inferred values of the DM components described in Section~\ref{sec:dm_model}, estimated for each FRB in our sample. From left to write, the columns show: the ID of a given FRB, spectroscopic redshift, mean and standard deviation of the \dmigm, inferred from the corresponding \texttt{ARGO} reconstructions, assuming $\figm=1.0$, the mean and standard deviation of the \dmhalo, assuming $\fgas=1.0$ and ${\rm r_{max}}/r_{200}=1.0$, the mean and standard deviation  of the FRB host' halo contribution assuming $\fgas = 1.0$, mean halo mass estimate for each FRB host based on the stellar masses from \citet{gordon2023}, combined contributions from the Milky Way's ISM and halo components, and the overall observed \dmfrb. All DM values are quoted in units of \pccm. }
\tablehead{
            \colhead{ FRB }      & 
            \colhead{ redshift } & 
            \colhead{ \dmargo  } &
            \colhead{ \dmhalo  } &
            \colhead{ ${\rm DM}_{\rm host}^{\rm halo}$ } &
            \colhead{ $ \langle \log_{10}\left( M_{\rm host}^{\rm halo}/ M_{\odot}\right)\rangle$ } &
            \colhead{ ${\rm DM_{MW}}$ } &
            \colhead{ \dmfrb } \\[-17pt]
            \colhead{} &
            \colhead{} &
            \colhead{} &
            \colhead{} & 
            \colhead{} &
            \colhead{} &
            \colhead{} &
            \colhead{} 
          }
\startdata
20211127A & 0.0469 & $61.6 \pm 32.0$  & $22.6\pm 13.2$  & $24.1\pm 5.5$ & 11.3 & $82.5$ & $234.83$ \\
20211212A & 0.0713 & $171.2\pm 106.4$ & $13.4\pm 15.7$  & $35.8\pm 8.1$ & 11.7 & $67.1$ & $206.00$ \\
20190608A & 0.1178 & $89.9 \pm 12.0$  & $30.5\pm 20.4$  & $46.1\pm 10.4$ & 12.1 & $78.1$ & $339.50$ \\
20200430A & 0.1608 & $116.1\pm 5.6$   & $120.0\pm 33.1$ & $24.2\pm 5.7$ & 11.2 & $67.0$ & $380.10$ \\
20191001A & 0.2340 & $466.3\pm 170.4$ & $360.7\pm 72.4$ & $58.1\pm 13.0$ & 12.3 & $84.7$ & $506.92$ \\
20190714A & 0.2365 & $326.5\pm 94.7$  & $321.4\pm 92.1$ & $37.4\pm 8.7$ & 11.8 & $78.0$ & $504.70$ \\
20180924B & 0.3212 & $329.2\pm 91.0$  & $11.5\pm 10.6$  & $44.0\pm 10.4$ & 11.9 & $81.9$ & $362.40$ \\
20200906A & 0.3688 & $559.9\pm 202.8$ & $37.4\pm 23.5$  & $45.8\pm 10.8$ & 11.9 & $75.9$ & $577.80$ 
\enddata
\end{deluxetable*}

\begin{figure}[t]
    \centering
    \includegraphics[width=\columnwidth]{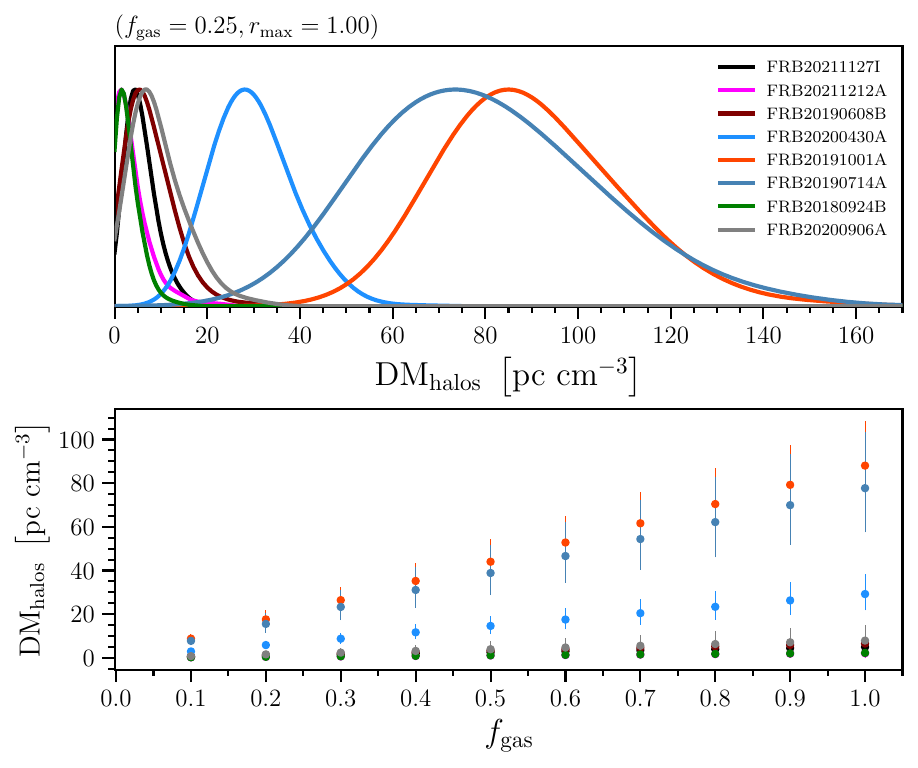}
    \caption{{\it Top:} KDE-smoothed PDF of the \dmhalo\ distributions estimated for each considered FRB field. {\it Bottom:} Estimated values of \dmhalo\ as a function of \fgas. Each marker is the $50^{\rm th}$ percentile of the corresponding distribution from the top panel, while the errorbars are given by $16^{\rm th}$ and $84^{\rm th}$ percentiles of the same distributions.}
    \label{fig:dmh_all}
\end{figure}

So far, we have considered only the average halo masses of the foreground halos. However, there are considerable uncertainties in the stellar mass estimation from the SED fitting technique.
These uncertainties, as well as the scatter in the stellar mass-halo mass relationship, propagate into the \dmhalos\ calculation. In order to take them into account, we introduce a relative random scatter $\sigma_{M_{\rm halo}}\left( \log_{10}\left[M_{\rm halo}/M_{\odot}\right] \right) = 0.3$~dex to the inferred halo masses. This is the typical uncertainty in converting \mstar\ to \mhalo\ for $\mstar=10^{10.5}M_\odot$ galaxies \citep{simha2021}. For each `narrow-field' galaxy catalog of each FRB sightline, we generate $N_{\rm real}=1000$ Monte Carlo realizations of the \dmhalos, where we randomly sample combinations of halo masses that would be consistent with the mean and scatter of the intervening galaxies' halo masses. For a given FRB sightline $i$, we compute the mean of these $N_{\rm real}=1000$ realizations, scaled by \fgas, and incorporate it in our model of the observed DM as

\begin{equation}
    \dmhalos = \fgas \frac{1}{N_{\rm real}} \sum_{j}^{N_{\rm real}} \sum_{k}^{N_{\rm halos}} \frac{ {\rm DM}_{{\rm halo},k}^{j} }{ 1+z_{{\rm halo},k} },
\end{equation}
where ${\rm DM}_{{\rm halo},k}$ of each $k$ foreground halo is corrected for the redshift dilation by a factor of $1 / \left( 1+z_{{\rm halo},k} \right)$. On the other hand, the standard deviation \shalo\ of the resulting Monte Carlo realizations of \dmhalos\ is then adopted for the likelihood calculation for the MCMC analysis (see Section~\ref{sec:like}).

We illustrate the resulting distribution of \dmhalos\ along each FRB sightline in the top panel of Figure~\ref{fig:dmh_all}. For illustrative purposes, we adopt a value of $f_{\rm gas}=0.25$. The bottom panel of Figure~\ref{fig:dmh_all}, meanwhile, shows the mean and standard deviation of these distributions as a function of the free parameter \fgas.

\subsection{The Hosts}
\label{sec:dm_host}

The contribution to the observed DM from the FRBs host galaxies remains a highly uncertain quantity. Previous studies have often used either a constant value or estimated it from an assumed probability density distribution \citep[e.g.][but see Bernales–Cortes et al. in prep. for an empirical estimation]{macquart2020, james2022a,james2022b}. In this work, we employ the following model for the ${\rm DM_{host}}$ 

\begin{equation}
\label{eq:dmhost}
    {\rm DM}_{{\rm host}, i} = \frac{ \dmhhalo  + \langle \dmhunkn \rangle}{1+z_{{\rm frb},i} },
\end{equation}
where \dmhhalo\ is the contribution from the CGM halo of the FRB host galaxy, and $\langle \dmhunkn \rangle$ is left as a free parameter in our inference model, describing the unknown contributions from the host ISM, FRB engine and its immediate surroundings.

In order to estimate \dmhhalo\ for a given FRB, we follow Section~\ref{sec:dm_halo} and use the mNFW model to calculate the halo density profile of the host galaxy, given published estimates of its stellar mass \citep{gordon2023}. To obtain the \dmhhalo\, we then integrate the resulting density profile along the path determined by the FRB impact parameter relative to the center of the host galaxy. In contrast to DM$_{\rm halos}$, here we integrate only half-way of the halo. Similar to the discussion in Section~\ref{sec:dm_halo}, we take into account the uncertainty in the measured masses of the host galaxies' halos by introducing a random scatter $\sigma\left(\log_{10}\left[ {M_{\rm host}^{\rm halo}}/ M_{\odot}\right] \right) = 0.30$~dex. For each FRB host, we then compute the mean and standard deviation, \shosth, of $N_{\rm real}=1000$ Monte Carlo realizations of the ${\rm DM_{host}^{halo}}$ given by

\begin{equation}
\dmhhalo = \fgas \frac{1}{N_{\rm real}} \sum_{j}^{N_{\rm real}} {\rm DM}_{\rm host}^{\rm halo} \left( M_{{\rm host},j}^{\rm halo}\right).
\end{equation}

We note that FRB~$20191001$A host galaxy is a special case among the DR1 sample because it has been found to reside within a galaxy group based on the methodology described in Section~\ref{sec:argo_grps} \citep[see also][]{bhandari2020}. As discussed in Section~\ref{sec:dm_halo}, we replace the information about the individual members with the corresponding properties of the group itself (mass, impact parameter, etc.) and use them to estimate the contribution which we assign to the \dmhalo{} (${\rm DM} = 166\pm50~\pccm$). On the other hand, to calculate the ${\rm DM_{host}^{halo}}$, we adopt the mean stellar/halo mass of the host galaxy itself, and found ${\rm DM_{host}^{halo}} = 58.1\pm 13.0~\pccm$ (assuming $\fgas=1.0$) in case of FRB~$20191001$A (see Table~\ref{tbl:tab_dms}). 
In other words, we include the contribution from both the host galaxy and its galaxy group, albeit in different DM terms.
This assumption should be tested with cosmological hydrodynamical simulations in future work.

FRB~$20191001$A is, to our knowledge, the fourth FRB known to reside within a galaxy group or cluster \citep{connor2023, gordon2023}, and it will be interesting to follow up such objects in future analyses to investigate possible insights this might have on the FRB host population.

\section{Parameter Inference}
\label{sec:mcmc}

In this section, we describe our inference algorithm and corresponding model parameters governing the evolution of DM components, discussed in Section~\ref{sec:dm_model}. We begin with the definition of the likelihood function, required by the MCMC algorithm later. 

\subsection{The Likelihood}
\label{sec:like}

We assume that the joint likelihood function $\mathcal{L}_{\rm frb}\left( {\rm DM}_{\rm frb} | \Theta \right)$ for $8$ FRBs in our sample (see Table~\ref{tbl:tab_dr1}) is well-described by a Gaussian

\begin{equation}
\label{eq:like}
\small
    \ln\mathcal{L}\left( {\rm DM_{FRB}} | \Theta \right) \propto -\frac{1}{2} \sum_i^{N_{\rm frb}} \left[ \frac{ \left( {\rm DM}_{{\rm model},i} \left( \Theta \right) - {\rm DM}_{{\rm FRB},i} \right)^2 }{\sigma_i^2} \right], 
\end{equation}
where $\Theta =\{ f_{\rm igm}, f_{\rm gas}, \langle \dmhunkn \rangle \}$ represents our model parameters, ${\rm DM}_{{\rm model},i}$ is the model dispersion measure, described in details in the previous section, and the model variance $\sigma^2_i$ is estimated by combining in quadrature uncertainties on the individual components of the total ${\rm DM}_{{\rm model},i}$, given by

\begin{align}
    \sigma_i^2 = \left(\sargo \right)^2 + \left(\shalo \right)^2 + 
    \left(\shost \right)^2 +\nonumber  \\ 
    \left(\shosth \right)^2 + \left(\smw \right)^2, 
\end{align}
where we omit the uncertainty on the observed \dmfrb{} because it is negligible in comparison to other considered uncertainties. For a given value of $\langle {\rm DM_{host}^{\rm unk}\rangle}$, we assume a log-normal distribution such that the corresponding variance $(\shost)^2$ is described by

\begin{equation}
    \left(\shost \right)^2 = \left( e^{\sigma_*^2} - 1\right) e^{\left(  2\mu + \sigma_*^2  \right)},
\end{equation}
where $\mu \equiv \langle \dmhunkn \rangle$, and we use the best-fit value $\sigma_* = 1.23$ from \citet{james2022a}\footnote{\citet{james2022a} reported the standard deviation of the log-normal ${\rm DM_{host}}$ distribution in $\log_{10}$ units ($\sigma=0.53$). We rescaled it by a factor of $1/\log_{10}e$ in order to convert to the natural logarithm units.}

\subsection{The Priors}
\label{sec:prior}

\begin{deluxetable*}{lcccc}[ht]
\label{tbl:priors}
\tablecaption{Bayesian Prior Combinations}
\tablehead{
            \colhead{} &
            \colhead{Flat $f_{\rm d}$} &
            \colhead{Flat $f_{\rm d}$ + Limited \fgas} &
            \colhead{Gaussian $f_{\rm d}$} &
            \colhead{Gaussian $f_{\rm d}$ + Limited \fgas} \\[-4pt]
            & \colhead{(Fiducial)} & & & \\[-15pt]
}    
\startdata
$\pi \left( \figm \right)$ & $\left( 0,1.00\right]$ & $\left( 0,1.00\right]$ & $\left( 0,1.00\right]$ & $\left( 0,1.00\right]$ \\
$\pi \left( \fgas \right)$ & $\left( 0,1.00\right]$ & $\left( 0,0.70\right]$ & $\left( 0,1.00\right]$ & $\left( 0,0.70\right]$ \\
$\pi \left( \ln{\dmhunkn} \right)$ & $\left[0,6.00\right]$ & $\left[0,6.00\right]$ & $\left[0,6.00\right]$ & $\left[0,6.00\right]$  \\
$\pi \left( f_{\rm d}\right)$ & $\left[0.75,0.95\right]$ & $\left[0.75,0.95\right]$ & $\mathcal{N}\left( \mu=0.86, \sigma=0.02 \right)$ & $\mathcal{N}\left( \mu=0.86, \sigma=0.02 \right)$
\enddata
\end{deluxetable*}

For the FLIMFLAM DR1 analysis, we will adopt several combinations of physically-motivated priors that we will describe here, summarized in Table~\ref{tbl:priors}. In all cases, we assume a flat linear prior on the fraction of cosmic baryons inside the IGM, i.e., $\pi \left( f_{\rm igm}\right)= \left( 0.0, 1.0\right]$. Similarly, also in all cases, we adopt a flat logarithmic prior on the unknown FRB host contribution $\pi\left(\ln{\langle \dmhunkn \rangle} \right)= \left[ 0.0,6.0 \right] $, consistent with the ranges explored by \citet{james2022b}.

For \fgas{}, we take as the default case a flat prior spanning all physical values between zero and unity: $\pi(\fgas)= \left( 0.0, 1.0\right]$. However, recently, \citet{khrykin2023} explored the evolution of the cosmic baryon fractions in and around simulated halos in the \texttt{Simba} suite of cosmological hydrodynamical simulations \citep{dave2019}. They found that the \fgas\ value depends significantly on both the considered halo mass range, and the exact feedback prescription used in the simulations. For the range of halo masses estimated for the foreground galaxies/groups in FLIMFLAM DR1 sample used in this work (see Figure~\ref{fig:hmasses}), results of \citet{khrykin2023} suggest a range of 
$\fgas = \left( 0, 0.70 \right]$ for all feedback prescriptions they considered. 
We therefore also consider an additional ``Limited \fgas" prior with $\pi(\fgas)= \left( 0.0, 0.7\right]$.

\begin{figure}[t]
    \centering
    \includegraphics[width=\columnwidth]{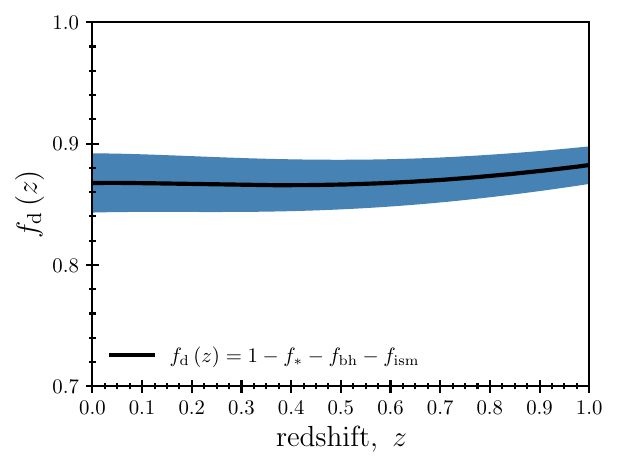}
    \caption{ Evolution of the cosmic diffuse baryon fraction $f_{\rm d}$ as a function of redshift. The shaded area illustrate the $2\sigma$ errorbars.}  
    \label{fig:fdz}
\end{figure}

We also include additional priors describing the total budget of baryons expected to be found in the diffuse states outside of individual galaxies,
\begin{align}\label{eq:fd}
    \fdnoz &\equiv \figm + \fcgm + \ficm \nonumber \\  &=1-f_*-f_{\rm bh}-f_{\rm ism},
\end{align}
where \ficm{} is the fraction of all cosmic baryons that reside 
within the intra-cluster media (ICM) of galaxy clusters with 
$M_{\rm halo} \gtrsim 10^{14}\,M_\odot$, and \fcgm{} is the fraction in lower-mass halos. 
Our free parameters are related to Equation~\ref{eq:fd} through
\figm{} explicitly,  and \fcgm{}, which is a function of \fgas. 
Meanwhile, \ficm{} is constrained by ICM gas mass fraction ($f_{\rm gas,icm}$) measurements using X-ray \citep[e.g.,][]{gonsalez2013, chiu2018} and Sunyaev-Zel'dovich \citep[e.g.,][]{degraaf2019} observations.

The overall \fdz{} can be estimated by the process of elimination: one estimates the total observed budget of stars, $f_*$, stellar remnants (neutron stars and black holes), $f_{\rm bh}$, and ISM in galaxies $f_{\rm ism}$, and then subtracts these from the total cosmic baryon to yield \fdnoz\ as seen in the second line of Equation~\ref{eq:fd}. 
To calculate \fdz, we follow the calculation presented in \cite{macquart2020} and encoded in 
the FRB repository\footnote{\url{https://github.com/FRBs/FRB}}.
The error in \fdz\ is dominated by the
systematic uncertainty in the stellar mass 
density \rhostar\ which depends on an assumed
stellar initial mass function (IMF).
The values reported by \cite{madau2014}, which are the defaults, assume the Salpeter IMF \citep{salpeter1955}. If we instead adopt the Chabrier IMF \citep{chabrier2003}, the \rhostar\ 
values decrease by 1.7 and \fdz\ increases at $z=0$ from $\approx 0.84$ to $\approx 0.9$. The resulting \fdz{} is shown as a function of redshift in Figure~\ref{fig:fdz}.
We use the \fdnoz{} evaluated at $\bar{z}=0.20$, the mean redshift probed by our sightlines, and implement it as a prior in two ways: 
(i) as a flat prior such that $\fdz=[0.75,0.95]$ (``Flat \fdnoz");
or (ii) as a Gaussian prior with a mean of $\mu=0.86$ and standard deviation 
$\sigma=0.02$ (``Gaussian \fdnoz").

Our analysis aims to constrain as one of our free parameters the quantity \fgas{}, which represents the fraction of baryons residing in the CGM on a per-halo basis. Therefore, in order to use the  prior in Equation~\ref{eq:fd}, we need to convert \fgas{} to \fcgm\, which is the fraction of cosmic baryons residing in all the halos in the Universe. For a range of the halo masses $\left[ M_1, M_2 \right]$, \fcgm\ is given by 
\begin{gather}
    \fcgm = \frac{1}{(\Omega_b/V) \int_V \bar{\rho}_{\rm m} \left( z\right) {\rm d}V} \times  \nonumber \\ 
      \int^{M_2}_{M_1} \left[ \int^{r_{\rm max}}_0 \! \fgas \Omega_b \rho_{\rm halo}\left(M_{\rm halo}, z, r\right) 4 \pi r^2 {\rm d}r\right] \times   \nonumber \\  \phi(M_{\rm halo}) \, {\rm d}\ln \frac{M_{\rm halo}}{M_{\odot}}, \label{eq:fcgm}
\end{gather}     
where $\bar{\rho}_{\rm m}\left( z\right)$ is the cosmic matter density at a given redshift, $\rho_{\rm halo}\left( z, r\right)$, the radial matter density profile of collapsed halos with mass $M_{\rm halo}$, and $\phi(M_{\rm halo})$ is the halo mass function (see more detailed discussion in \citealp{khrykin2023}). 
The conversion between $f_{\rm gas,icm}$ and \ficm{} is similar, except that the mass range is specifically set to $M_{\rm halo} \geq 10^{14}\,M_\odot$.
In the case of the ICM, we assume a fixed $f_{\rm gas,icm}=0.8$ with a Gaussian standard deviation of $\sigma=0.1$, which is consistent with current measurements of gas in galaxy clusters \citep{gonsalez2013, chiu2018}.
Adopting the halo mass function from the Aemulus package \citep{mcclintock2019} and assuming the mean redshift of our sample $\bar{z}\!\!\simeq\! 0.20$, and $\rmax = 1.0 \times r_{200}$, we pre-compute a lookup reference table between \fgas\ and $\fcgm$ values (and equivalently, $f_{\rm gas,icm}$ and \ficm{}) using Equation~(\ref{eq:fcgm}). 

While the definition of \fcgm{} in Equation~\ref{eq:fd} is supposed to span the entire range of non-cluster halo masses in the Universe ($M<10^{14}\,M_\odot$), the FLIMFLAM DR1 ``narrow-field" foreground data does not fully cover this entire mass range (Figure~\ref{fig:hmasses}).
We therefore further split \fcgm{} into two terms, 
$\fcgm=\fcgmff+\fcgmother$. The \fcgmff{} represents the halo mass range sampled by our data, for which \fgas{} is a free parameter. For halo masses not represented by our data, \fcgmother{} is the unknown contribution to the cosmic budget. 
We assume that the \fgas{} for these halos can span $0 < \fgas \leq 1$ with uniform probability.

Consequently, at each MCMC step, the proposed value of \fgas\ is then converted to \fcgmff, while random realizations of \ficm{} and \fcgmother{} are drawn, making the aforementioned assumptions for their respective \fgas{}.
These terms are then compared to \fdnoz{} in Equation~(\ref{eq:fd})
to apply the aforementioned prior. 

Note that, in general, \fgas{} is expected to be a function of halo mass \citep{ayromlou2023,khrykin2023}. Future analyses should adopt more sophisticated parametrization, but for FLIMFLAM DR1 the simplified assumption described above should suffice given the limited data.

\begin{deluxetable*}{lcccc}[ht]
\label{tbl:tab_results_priors}
\tablecaption{Results of the MCMC analysis for different set of priors $\pi\left( \Theta \right)$. The last row shows the values of the total host contribution (both, from the corresponding halo and the unknown contribution from the host' ISM and FRB engine), averaged over 8 FRB sightlines used in this work.}
\tablehead{
            \colhead{} &
            \colhead{Flat $f_{\rm d}$} &
            \colhead{Flat $f_{\rm d}$ + Limited \fgas} &
            \colhead{Gaussian $f_{\rm d}$} &
            \colhead{Gaussian $f_{\rm d}$ + Limited \fgas} \\[-4pt]
             & \colhead{(Fiducial)} & & & \\[-15pt]
 }
\startdata
$\figm$ & $0.59^{+0.11}_{-0.10}$ & $0.63^{+0.09}_{-0.07}$ & $0.59^{+0.11}_{-0.10}$ & $0.64^{+0.09}_{-0.07}$ \\[4pt]
$\fgas$ & $0.55^{+0.26}_{-0.29}$ & $0.44^{+0.18}_{-0.24}$ & $0.55^{+0.26}_{-0.29}$ & $0.44^{+0.18}_{-0.24}$ \\[4pt]
$\langle \dmhunkn \rangle$ & $69^{+28}_{-19}~{\rm pc~cm^{-3}}$ & $70^{+29}_{-18}~{\rm pc~cm^{-3}}$ & $69^{+29}_{-19}~{\rm pc~cm^{-3}}$ & $70^{+29}_{-19}~{\rm pc~cm^{-3}}$ \\[4pt]
$\langle {\rm DM_{host}} \rangle$ & $ 90^{+29}_{-19}~{\rm pc~cm^{-3}}$ & $86^{+28}_{-18}~{\rm pc~cm^{-3}}$ & $90^{+31}_{-20}~{\rm pc~cm^{-3}}$ & $87^{+29}_{-19}~{\rm pc~cm^{-3}}$\\[4pt]
\enddata
\end{deluxetable*} 

\begin{figure*}
    \centering
    \includegraphics[width=\textwidth]{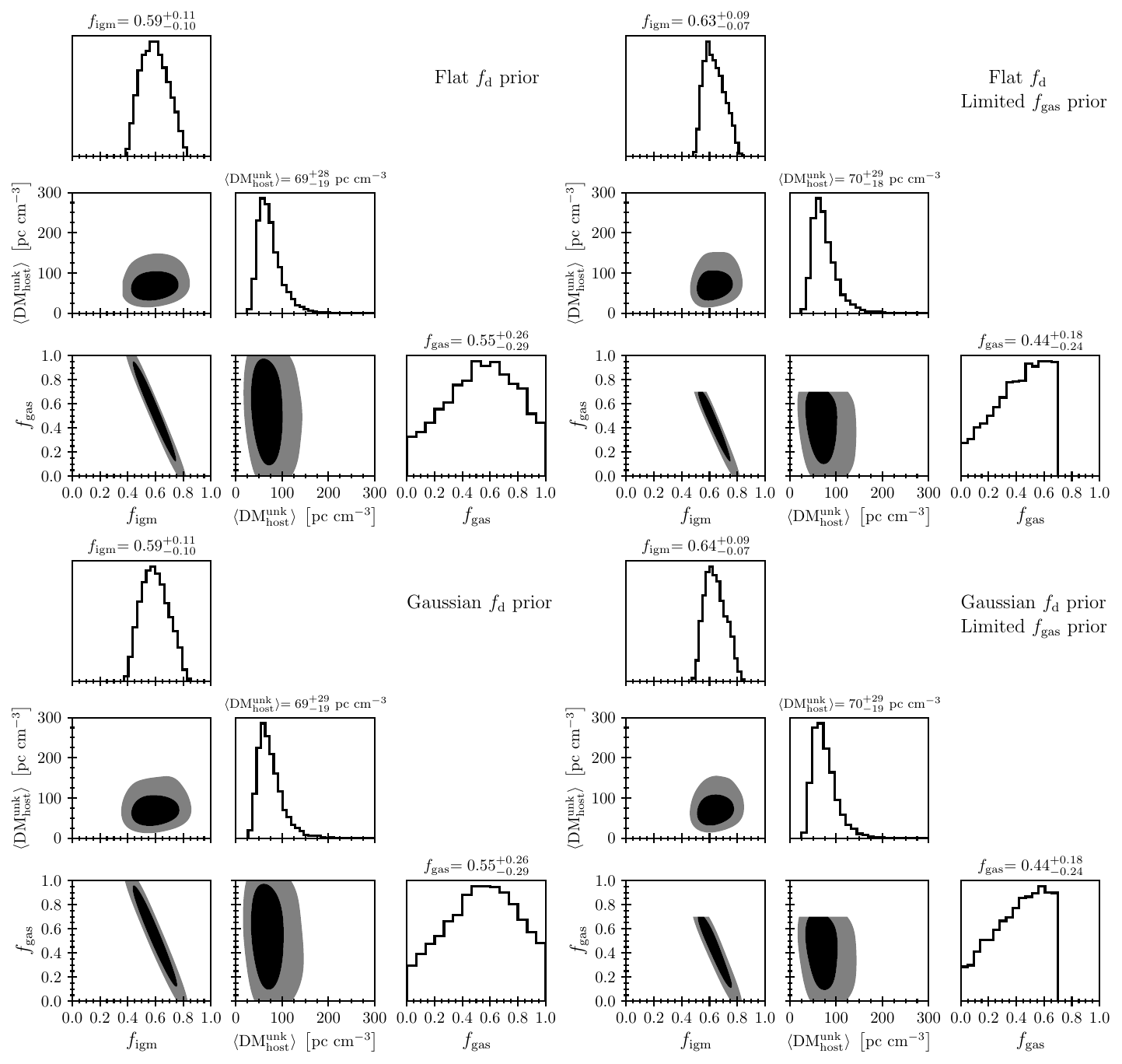}
    \caption{ Results of the MCMC inference on the observed sample of FRB in Table~\ref{tbl:tab_dr1}. The panels show inferred $2$D contours and marginalized 1D posterior probabilities of the model parameters. The black and gray contours correspond to the $68\%$ and $95\%$ confidence regions, respectively. Each triangle plot corresponds to a specific choice of priors. See text for details.}  
    \label{fig:mcmc_data}
\end{figure*}

\begin{figure*}
    \centering
    \includegraphics[width=\textwidth]{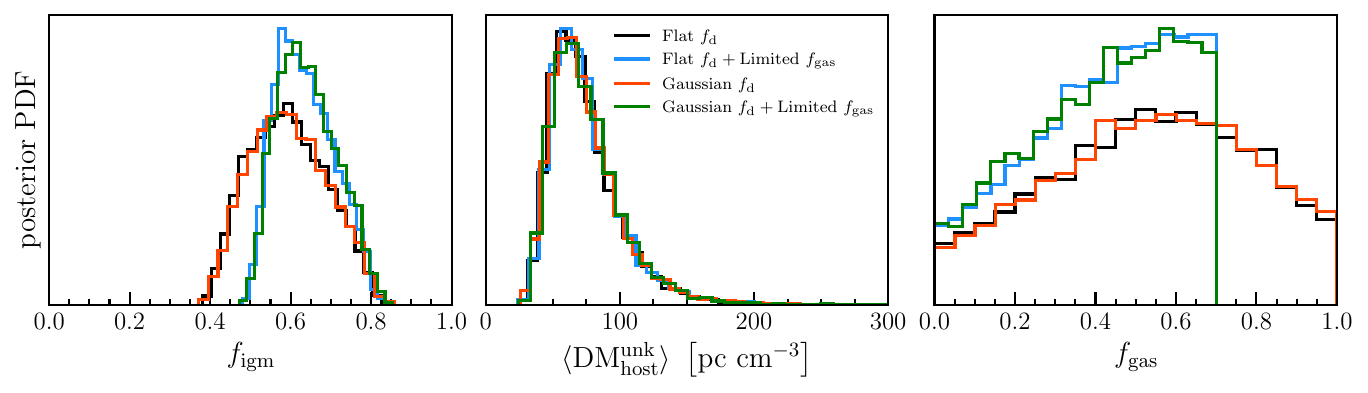}
    \caption{ 1D marginalized posterior distributions for each model parameter, estimated using different combinations of the priors. The black curve shows the results corresponding to the fiducial set of priors, discussed in Section~\ref{sec:prior}, whereas the other curves correspond to the modified priors on \fgas\ and $f_{\rm d}\left(z\right)$.  }  
    \label{fig:1d_post_priors}
\end{figure*}

\section{Results}
\label{sec:results}

Given the expression for the joint likelihood in Equation~(\ref{eq:like}), and the choice of priors described in Section~\ref{sec:prior}, we now proceed to sample the aforementioned likelihood function using the MCMC algorithm in order to estimate the posterior probability distributions for the model parameters \figm, \fgas, and $\langle \dmhunkn \rangle$. We adopt the publically available affine-invariant MCMC sampling algorithm \texttt{EMCEE} \citep{fm2013}.

The results of the MCMC inference are shown in Figure~\ref{fig:mcmc_data}, where the 2D contours illustrate the 95\% (gray) and 68\% (black) confidence intervals, respectively. 1D marginalized posterior probability distributions for each model parameter are also shown by the corresponding KDE histograms, while the values are tabulated in Table~\ref{tbl:tab_results_priors}.

For the purpose of the discussion, we adopt the `Flat \fdnoz{}'-only case as the fiducial prior (c.f.\ Table~\ref{tbl:priors}). In this case, we estimate the IGM baryon fraction to be $\figm = 0.59^{+0.11}_{-0.10}$, and the unknown DM contribution from the host ISM and FRB engine to be $\langle \dmhunkn \rangle= 69^{+28}_{-19}~\pccm$. 

On the other hand, results presented in Figure~\ref{fig:mcmc_data}, indicate that the current sample of the FRBs used in this work has only limited sensitivity to the \fgas\ value: we find $\fgas = 0.55^{+0.26}_{-0.29}$, and within the 95th confidence level the 1D and 2D contours span the entire range of allowed \fgas{} values (Figure~\ref{fig:mcmc_data}).
Applying Equation~(\ref{eq:fcgm}), this implies that $\fcgmff = 0.20^{+0.10}_{-0.11}$ of the baryons in the Universe exists as CGM gas surrounding halos with $10^{11}\,M_\odot \lesssim \mhalo \lesssim 10^{13}\,M_\odot$.
There is a degeneracy between \figm{} and \fgas{} primarily set by the prior constraints on the total amount of diffuse baryons, \fdnoz, residing outside of galaxies (see Section~\ref{sec:prior}).

In comparison with the fiducial Flat \fdnoz{} prior, changing to the Gaussian \fdnoz{} case leads to negligible changes to the resulting parameter constraints and errors at the $\sim 1\%$ level. However, adopting a more limited range of $\fgas=(0.00,0.70]$ as suggested by hydrodynamical simulations \citep{khrykin2023} leads to more noticeable differences.
As expected, \fgas{} is reduced from $\fgas =0.55^{+0.26}_{-0.29}$ in the fiducial case to $\fgas =0.44^{+0.18}_{-0.24}$, corresponding to a CGM fraction of $\fcgmff = 0.16^{+0.07}_{-0.09}$ for our halo mass range. With the reduction of \fgas{} from the Limited-\fgas{} prior, \figm{} also increases along the \figm-\fgas{} degeneracy direction to $\figm=0.63^{+0.09}_{-0.07}$. The difference in the recovered 1D marginalized posterior distributions of each parameter in different MCMC runs is shown in Figure~\ref{fig:1d_post_priors}.

Our constraints of $\figm\approx 0.59-0.64$ are consistent with an IGM that has experienced no galaxy feedback of any kind, which should yield $\figm\approx 0.59$ according to the cosmological hydrodynamical simulations analyzed by \citet{khrykin2023}. However, irrespective of the chosen priors, given the uncertainties, our resulting constraints on \figm{} are also consistent with an IGM that has experienced stellar feedback ($\figm \simeq 0.70$ per \citealt{khrykin2023}). On the other hand, \citet{khrykin2023} showed that AGN jet feedback ejects significant amounts of baryons out into the IGM, resulting in higher IGM baryon fractions in the relevant simulations: $\figm \approx 0.85-0.87$. Therefore, our constraints mildly disfavor the AGN jet feedback scenario at the 2-sigma level. 

For all the prior combinations we have considered here, however, the resulting constraints on the `unknown' host ISM and host engine contribution to \dmhost{} remains consistently at $\langle \dmhunkn \rangle \approx 70\,\pccm$. This indicates that our data are successful at separating out the \dmhost{} component from the total observed DM, even though we only weakly constrain the further separation of the \dmigm{} and $\mathrm{DM_{halos}}$ components.

As mentioned in Section~\ref{sec:dm_host}, the contribution of the FRB host galaxy ${\rm DM_{host}}$ to the observed DM is driven by two components: the host' extended CGM halo ${\rm DM_{host}^{halo}}$ and by the hitherto unconstrained input from the host' stellar/ISM environment and the FRB progenitor itself or its immediate surroundings, $\langle \dmhunkn \rangle$. Following Equation~(\ref{eq:dmhost}), we can estimate the mean total host contribution  $\langle {\rm DM_{host}} \rangle$, averaged over our $8$ FRB sightlines, which is given by

\begin{equation}
\label{eq:dmhost_mean}
    \langle {\rm DM_{host}} \rangle = \frac{1}{N_{\rm frb}} \sum_{i}^{N_{\rm frb}} \fgas  \dmhhalo + \langle \dmhunkn \rangle,
\end{equation}
where we omitted the factor of $1\slash \left( 1+ z_{\rm frb}\right)$ in Equation~(\ref{eq:dmhost_mean}) since it is already taken into account previously (see discussion in Sections~\ref{sec:dm_host}). 

To fully capture the covariance between \fgas{} and $\langle \dmhunkn{} \rangle$ estimated by the MCMC algorithm in Section~\ref{sec:mcmc}, we sample the posterior pairs of these parameters from the corresponding MCMC chains. Moreover, for each pair of \fgas{} and $\langle \dmhunkn{} \rangle$ drawn from the MCMC chain, we additionally randomly choose a value of the host galaxy halo mass from the corresponding Gaussian distribution (see Section~\ref{sec:dm_host}) and calculate the \dmhhalo{} in Equation~(\ref{eq:dmhost_mean}). This results in $\approx 10000$ realization of $\langle {\rm DM_{host}} \rangle$ per each MCMC run. Finally, we estimate the corresponding $16^{\rm}$, $50^{\rm}$, and $84^{\rm}$ percentiles of these distributions. We report the resulting average host galaxy contribution to the observed DM and the corresponding uncertainties in the last row of Table~\ref{tbl:tab_results_priors}.

In all considered prior combinations, we find consistent mean values of $\langle {\rm DM_{host}}\rangle \approx 86 - 90~\pccm$, and comparable uncertainties. This value is somewhat lower than reported by \citet{james2022a} (and subsequently \citealp{baptista2023}). In their analysis of nearly $70$ FRBs, they found $\langle {\rm DM_{host}}\rangle \approx 130^{+66}_{-48}~\pccm$, and a log-normal scatter of $\log_{10}\sigma \simeq 0.50$. 

Although our $\langle {\rm DM_{host}}\rangle$ is formally consistent with their result, we believe our lower value can probably be explained by the fact that we have explicitly rejected from our sample FRBs believed to have large $\langle {\rm DM_{host}}\rangle$, e.g.\ FRB20210117A \citep{simha2023} and FRB20190520B \citep{kglee2023}. \citet{james2022a}, on the other hand, did not make use of any foreground information, and therefore such excess-DM objects would have been included. In the future, it is possible that other measured FRB quantities, such as the scattering time (e.g., \citealt{cordes2022}) or H$\alpha$ emission measure \citep{tendulkar2017}, could be used as priors for the $\dmhunkn$ contribution from individual FRBs in the analysis sample (Bernales-Cortes et al., in prep).

\section{Conclusions}
\label{sec:end}

In this work, we presented the analysis of the first data release of the FLIMFLAM survey, aimed at revealing the distribution of the comic baryons within the diffuse IGM as well as the CGM gas of galaxy halos.

We have conducted an extensive observational campaign to collect wide-field spectroscopic information on galaxies in the foreground of $8$ localized FRBs at $z \lesssim 0.4$ (with a mean redshift of $\bar{z}\simeq 0.20$ probed by the sample). This information has been used in the state-of-the-art \texttt{ARGO} Bayesian statistical algorithm to reconstruct the matter density field along each FRB sightline, permitting models for the contribution of the diffuse IGM gas to the observed FRB DMs. We have also collected narrow-field spectroscopic data about the foreground halos intersected by the FRB. This allowed us to estimate their contribution to the observed FRB \dmfrb. 

We have then utilized a Bayesian algorithm to statistically compare the observed FRB DMs to the theoretical model predictions. The main results of our work are as follows:

\begin{itemize}
    \item Assuming the fiducial set of flat priors on the model parameters, we measure the fraction of baryons residing in the IGM to be $\figm = 0.59^{+0.11}_{-0.10}$, and the corresponding $\fcgm = 0.20^{+0.10}_{-0.11}$ of baryons in the Universe inside the CGM of $10^{10}M_{\odot} \lesssim M_{\rm halo} \lesssim 10^{13}M_{\odot}$ halos. Imposing a more strict prior on \fgas{}, motivated by the hydrodynamical simulations, modifies this partition of cosmic baryons to $\figm = 0.63^{+0.09}_{-0.07}$ and $\fcgm = 0.16^{+0.07}_{-0.09}$, respectively. 
    
    \item Our results on \figm{} so far appear to be consistent with the predictions of recent hydrodynamical simulations for the IGM gas. However, given the large uncertainties in this preliminary sample, we cannot rule out any of the feedback models analyzed by \citet{khrykin2023}. 
    \item Based on our sample, we find that the host galaxies on average contribute $\langle {\rm DM_{host}} \rangle = 90^{+29}_{-19}$~\pccm\ to the observed \dmfrb{} (fiducial priors), while the major part of it is coming from the `unknown' host ISM and/or FRB engine, adding on average $\langle \dmhunkn \rangle = 69^{+28}_{-19}$~\pccm. This is the first analysis to attempt to separate out these components.
\end{itemize}

The information, encoded in the foreground structures traversed by the FRB, is paramount for placing high-precision constrains on cosmological and astrophysical parameters \citep[e.g.][]{simha2020,simha2023,kglee2023}. The future complete FLIMFLAM sample of $N\simeq 20$ FRBs will provide a more robust estimate of both, the distribution of baryons and the involvement of various feedback mechanisms in shaping it. Beyond FLIMFLAM, increasing localization efforts by the  Commensal Real-Time ASKAP Fast Transient \citep[CRAFT;][]{macquart2010}, the MeerKAT TRAnsients and Pulsars \citep[MeerTRAP;][]{sanidas2018}, and the Deep Synoptic Array \citep[DSA;][]{kocz2019}, coupled with observational campaigns on 4m-class multiplexed instruments such as DESI \citep{levi2013} will achieve detailed measurements of the cosmic baryons distribution in the low-redshift Universe ($z\lesssim 0.3$) on samples of $N \gtrsim 100$ FRBs.

\section*{Acknowledgements} 
\label{sec:acknowledgements}
We thank Elmo Tempel for kindly providing his group-finding software. Kavli IPMU is supported by World Premier International Research Center Initiative (WPI), MEXT, Japan. This work was performed in part at the Center for Data-Driven Discovery, Kavli IPMU (WPI). I.S.K. and N.T. would like to acknowledge the support received by the Joint Committee ESO-Government of Chile grant ORP 40/2022. R.M.S. acknowledges support through the Australian Research Council Future Fellowship FT190100155. R.M.S. and A.T.D. acknowledge support through the Australian Research Council Discovery Project DP220102305. J.X.P., S.S., A.C.G. and the Fong Group at Northwestern acknowledges support by the National Science Foundation under grant Nos. AST-1909358, AST-2308182 and CAREER grant No. AST-2047919. A.C.G acknowledges support from NSF grants AST-1911140, AST-1910471 and AST-2206490 as a member of the Fast and Fortunate for FRB Follow-up team. L.M. acknowledges the receipt of an MQ-RES scholarship from Macquarie University. M.G. is supported by the Australian Government through the Australian Research Council's Discovery Projects funding scheme (DP210102103).

Parts of this research were supported by the Australian Research Council Centre of Excellence for All Sky Astrophysics in 3 Dimensions (ASTRO 3D), through project No. CE170100013. Based in part on data acquired at the Anglo-Australian Telescope, under programs A/2020B/04, O/2021A/3001, A/2021A/13, A/2021B/9, and A/2022A/9. We
acknowledge the traditional custodians of the land on which the AAT stands, the Gamilaraay people, and pay our respects to elders past and present.

The data presented herein were obtained at the W.M. Keck Observatory (PIDs: U051, U100 and U162), which is operated as a scientific partnership among the California Institute of Technology, the University of California and the National Aeronautics and Space Administration. The Observatory was made possible by the generous financial support of the W.M. Keck Foundation. We also wish to recognize and acknowledge the very significant cultural role and reverence that the summit of Maunakea has always had within the indigenous Hawai’ian community. We are most fortunate to have the opportunity to conduct observations from this mountain.

Based on observations collected at the European Southern Observatory under ESO programmes: 
2102.A-5005(A), 0104.A-0411(A), 
105.20HG.001, 
110.241Y.001, and 
110.241Y.002. 

Based on observations obtained at the international Gemini Observatory (PID: GS-2022B-Q-137), a program of NSF NOIRLab, which is managed by the Association of Universities for Research in Astronomy (AURA) under a cooperative agreement with the U.S. National Science Foundation on behalf of the Gemini Observatory partnership: the U.S. National Science Foundation (United States), National Research Council (Canada), Agencia Nacional de Investigaci\'{o}n y Desarrollo (Chile), Ministerio de Ciencia, Tecnolog\'{i}a e Innovaci\'{o}n (Argentina), Minist\'{e}rio da Ci\^{e}ncia, Tecnologia, Inova\c{c}\~{o}es e Comunica\c{c}\~{o}es (Brazil), and Korea Astronomy and Space Science Institute (Republic of Korea).

This scientific work uses data obtained from Inyarrimanha Ilgari Bundara / the Murchison Radio-astronomy Observatory. We acknowledge the Wajarri Yamaji People as the Traditional Owners and native title holders of the Observatory site. CSIRO’s ASKAP radio telescope is part of the Australia Telescope National Facility (https://ror.org/05qajvd42). Operation of ASKAP is funded by the Australian Government with support from the National Collaborative Research Infrastructure Strategy. ASKAP uses the resources of the Pawsey Supercomputing Research Centre. Establishment of ASKAP, Inyarrimanha Ilgari Bundara, the CSIRO Murchison Radio-astronomy Observatory and the Pawsey Supercomputing Research Centre are initiatives of the Australian Government, with support from the Government of Western Australia and the Science and Industry Endowment Fund.

This research has made use of data obtained from the SuperCOSMOS Science Archive, prepared and hosted by the Wide Field Astronomy Unit, Institute for Astronomy, University of Edinburgh, which is funded by the UK Science and Technology Facilities Council. 

This publication makes use of data products from the Two Micron All Sky Survey, which is a joint project of the University of Massachusetts and the Infrared Processing and Analysis Center/California Institute of Technology, funded by the National Aeronautics and Space Administration and the National Science Foundation. 

This research has made use of the NASA/IPAC Extragalactic Database (NED) which is operated by the Jet Propulsion Laboratory, California Institute of Technology, under contract with the National Aeronautics and Space Administration. This publication makes use of data products from the Wide-field Infrared Survey Explorer, which is a joint project of the University of California, Los Angeles, and the Jet Propulsion Laboratory/California Institute of Technology, funded by the National Aeronautics and Space Administration.

%

\vspace{5mm}
\facilities{AAT(2dF-AAOmega), VLT(MUSE), Keck(DEIMOS, LRIS, KCWI)}


\software{MARZ \citep{hinton2016}, Astropy \citep{astropy2022}, Numpy \citep{harris2020}, EMCEE \citep{fm2013}, CIGALE \citep{baptista2023}, Matplotlib \citep{hunter2007} }

\bibliography{cobra_dr1}{}
\bibliographystyle{aasjournal}
\end{document}